\documentclass{jfm}
\usepackage{graphicx}
\usepackage{epstopdf, epsfig}
\usepackage{enumerate}

\newcommand{\bs}{\boldsymbol}

\usepackage{mathtools}
\usepackage{graphicx, subfigure}
\usepackage{chemarrow}
\usepackage{dsfont}

\usepackage{mhchem}

\usepackage{xcolor}
\colorlet{shadecolor}{orange!15}

\newcommand{\beq}{\begin{equation}}
\newcommand{\eeq}{\end{equation}}
\newcommand{\beqa}{\begin{eqnarray}}
\newcommand{\eeqa}{\end{eqnarray}}
\newcommand{\bem}{\begin{math}}
\newcommand{\eem}{\end{math}}
\newcommand{\rar}{{\rightarrow}}

\newcommand{\nvec}{\hat{\bs n}}

\newcommand{\bff}{{\bf f}}

\newcommand{\bfv}{{\bf v}}
\newcommand{\bfu}{{\bf u}}

\newcommand{\bfJ}{{\bf J}}

\newcommand{\gammapert}[1]{{#1}^{(\gamma)} \ }
\newcommand{\aver}[1]{\left\langle {#1}\right\rangle}

\def\tl#1{\textcolor{black}{#1}}

\def\tbl#1{\textcolor{black}{#1}}
\def\yhy#1{\textcolor{black}{#1}}

\shorttitle{Self-phoresis of a Pt-insulator Janus swimmer}
\shortauthor{Y. Ibrahim, R. Golestanian and T. B. Liverpool}

\title{Multiple phoretic mechanisms in the self-propulsion of a Pt-insulator Janus swimmer}

\author{Yahaya Ibrahim\aff{1},
  Ramin Golestanian\aff{2}
 \and Tanniemola B. Liverpool\aff{1,3}}

\affiliation{\aff{1}School of Mathematics, University of Bristol , University Walk, Bristol BS8 1TW, UK
\aff{2}Rudolf Peierls Centre for Theoretical Physics, 1 Keble Road, Oxford, OX1 3NP, UK
\aff{3} BrisSynBio, Tyndall Avenue, Bristol, BS8 1TQ, UK }

\begin{document}

\maketitle
%\newcommand{\bs}{\boldsymbol}
%
%\usepackage{mathtools}
%\usepackage{graphicx, subfigure}
%\usepackage{chemarrow}
%\usepackage{dsfont}
%
%
%\usepackage{xcolor}
%\colorlet{shadecolor}{orange!15}
%
%
%\title[Self-phoresis of a Pt-insulator Janus swimmers.]{Self-phoresis of a Pt-insulator Janus swimmer.}
%
%\author[Y. Ibrahim, R. Golestanian and T. B. Liverpool]%
%{Y. Ibrahim$^1$%
%  \thanks{yahaya.ibrahim@bristol.ac.uk; ramin.golestanian@oxford.ac.uk; t.liverpool@bristol.ac.uk}, 
%R. Golestanian$^2$,  and T. B. Liverpool$^1$ }
%
%% NOTE: A full address must be provided: department, university/institution, town/city, zipcode/postcode, country.
%\affiliation{$^1$School of Mathematics, University of Bristol - Clifton, Bristol BS8 1TW, UK\\[\affilskip]
%$^2$Rudolf Peierls Centre for Theoretical Physics, 1 Keble Road, Oxford, OX1 3NP, UK}
%
%\pubyear{2014}
%\volume{650}
%\pagerange{119--126}
%% Do not enter received and revised dates. These will be entered by the editorial office.
%\date{\today}
%%\setcounter{page}{1}
%
%\begin{document}
%
%
%%%%%%%%%%%%%%%%%%%%%%
%%% rate-limiting kinetics consideration 
%%%%%%%%%%%%%%%%%%%%%%%%
%
%
%
%\maketitle

\begin{abstract}
We present a detailed theoretical study which demonstrates that electrokinetic effects can also play a role in the motion of \tl{metallic-insulator} spherical Janus particles. Essential to our analysis is the identification of the fact that the reaction rates depend on Pt-coating thickness and that the thickness of coating varies from pole to equator of the coated hemisphere. We find that their motion is due to a combination of neutral and ionic diffusiophoretic as well as electrophoretic effects whose interplay can be changed by varying the ionic properties of the fluid. This has great potential significance for optimising performance of designed synthetic swimmers. 
\\
\textbf{Key ideas:} (1.) non-uniform reaction rates due to Pt-coating thickness variation, (2.) charged intermediates in the $\text{H}_2\text{O}_2$ catalysis by the Platinum.% \textbf{Hypothesis:} The coupled non-uniform reaction rates over the surface and charge intermediates generates electric current in the Pt cap from slow reaction sites to fast reaction sites. Hence proton current in the fluid compelets the cycle. The proton current drags the fluid along, resulting in the apparent slip velocity. 
\end{abstract}

\section{Introduction}

\begin{figure}
\begin{center}
\includegraphics[scale=.35]{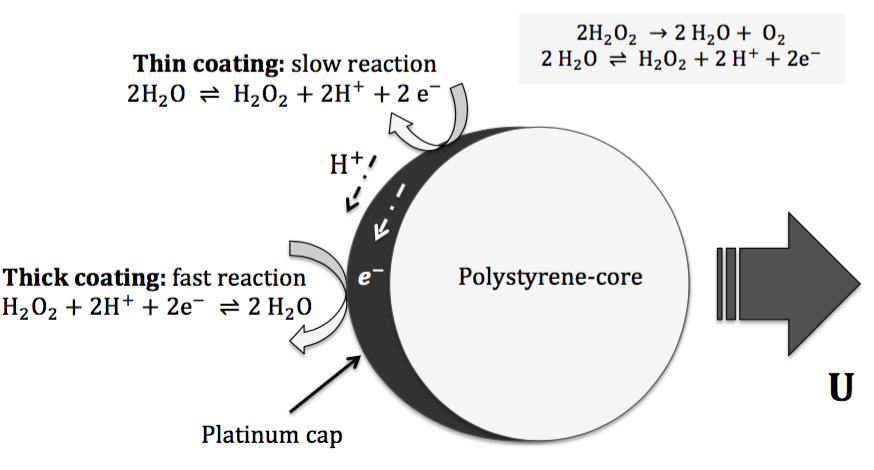}
\caption{\textcolor{black}{Cross-section of a schematic swimmer showing the variation of the thickness of  the Pt-coating, the directions of the currents and swimming direction.}}
\label{fig:1}
\end{center}
\end{figure}

In recent years there has been a flurry of activity in developing micro- and nanoscale self-propelling devices that are engineered to produce enhanced motion within a fluid environment~\citep{Kapral2013}. They are of interest for a number of reasons, including the potential to perform transport tasks~\citep{Patra2013}, and exhibit new emergent phenomena~\citep{Marchetti2013,Volpe2011,Theurkauff2012,Palacci2013,Kummel2013b,Bricard2013}. A variety of subtly different methods, all based on the catalytic decomposition of dissolved fuel molecules, have been shown to produce autonomous motion, or swimming.
%Many of the examples rely on the same catalytic reaction, the breakdown of hydrogen peroxide (H$_2$O$_2$) into water (H$_2$O) and oxygen (O$_2$) with metallic platinum (Pt) as catalyst. %In addition to the shape of the device, the spatial distribution of platinum on the surface interacting with the fuel solution plays an integral part in determining its motion.
Commonly studied systems are catalytic bimetallic rod shaped devices~\citep{Kline2005} and \tl{metallic-insulator} spherical Janus particles that are half-coated with catalyst (e.g. Platinum) for a non-equlibrium reaction (e.g. the decomposition of Hydrogen Peroxide)~\citep{Howse2007} [see Figure \ref{fig:1} (a)]. The propulsion mechanism is thought to be phoretic in nature~\citep{Anderson1989,Golestanian2007}, \tbl{but many specific details, such as which type of phoretic mechanism drive propulsion, remain the subject of debate~\citep{Golestanian2007,Gibbs2009a,Brady2011,Moran2011a}.} %More and better experimental data are required to test the proposed mechanisms.
A fundamental understanding of the mechanisms is key for developing the knowledge of how to use and control them in applications, and how to build up a picture of the collective behaviour through implementation of realistic interactions between catalytic colloids. 

For bimetallic swimmers, a plausible proposal is that the two metallic segments, usually platinum and gold, electrochemically reduce the dissolved fuel, in a process that results in electron transfer across the rod~\citep{Paxton2005,Kagan2009}. This together with proton movement in the solution \citep{Farniya2013c} {\em and} the interaction between the resulting self-generated electric field and the charge density on the rod produces (self-electrophoretic) motion~\citep{Moran2011a}. The direction of travel and swimming speed for arbitrary pairs of metals are well understood in the context of this mechanism~\citep{Wang2006}, as well as the link between fuel concentration and velocity~\citep{Sabass2012b}. For Pt-insulator Janus particles, the absence of conduction between the two hemispheres suggests a mechanism independent of electrokinetics. Hence, a natural first proposal is that a self-generated gradient of product and reactants can lead to motion via self-diffusiophoresis \citep{Golestanian2005}, provided the colloid is sufficiently small \citep{Gibbs2009a}. A number of predictions have been made based on this mechanism \citep{Golestanian2005,ruckner2007,Sabass2010,Valadares2010a,Popescu2009b,Brady2011,Sharifi-Mood2013} which have to date shown good agreement with the experimental dependency of swimming velocity on the size of the colloid~\citep{Ebbens2012}, and fuel concentration~\citep{Howse2007}. It would thus appear that a key difference between the bimetallic and \tl{metallic-insulator} Janus particles is that the motility in the latter system does not {\em require} conduction or electrostatic effects. However recent experiments have raised the possibility that this assumption might not be completely correct~\citep{Ebbens2014,Brown2014,Das2015}.

\begin{figure}
\begin{center}
\includegraphics[scale=.35]{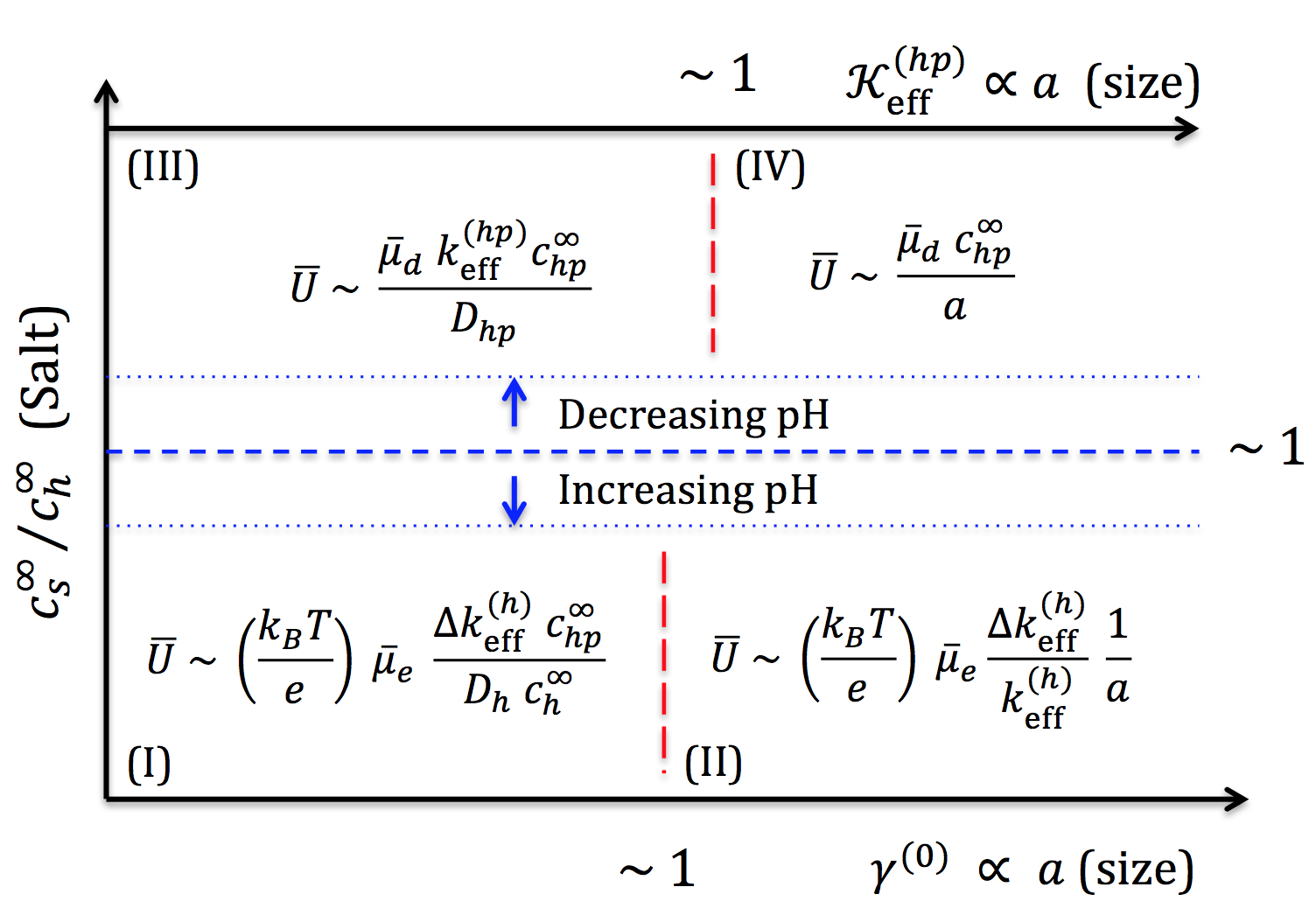}
\caption{\tl{Asymptotic regimes of the swimmer  propulsion speed $\bar{U}$. The dimensionless parameters on the horizontal axes are $\gamma^{(0)} = k_{\text{eff}}^{(h)} c_{hp}^{\infty}a/D_h\sum_{i\in \text{ions}} c_i^{\infty}$ and $\mathcal{K}_{\text{eff}}^{(hp)} = k_{\text{eff}}^{(hp)}a/D_{hp}$ where $D_i$ and $c_i^{\infty}$  are respectively the diffusion coefficient and bulk concentration of chemical specie $i$. $k_{\text{eff}}^{(i)}$ is the average rate of production/consumption of specie $i$ on the catalytic coated hemisphere.  $\Delta k_{\text{eff}}^{(i)}$ is the difference in the reaction rate between the equator where the coating is thinnest and the pole where the coating is thickest. $\bar{\mu}_e$ and $\bar{\mu}_d$ are electrophoretic and diffusiophoretic mobilities respectively.} } \label{fig:phase:diag}
\end{center}
\end{figure}

Here we present a detailed  theoretical study which demonstrates however that electrokinetic effects \citep{Pagonabarraga2010} can also play a role in the motion of \tl{metallic-insulator} spherical Janus particles expanding on our previous analyses briefly presented in ~\citep{Ebbens2014}. We find that their motion is due to a combination of neutral and ionic diffusiophoretic as well as electrophoretic effects whose interplay can be changed by varying the ionic properties of the fluid \yhy{(see Fig. \ref{fig:phase:diag})}. This has great potential significance as the effect on the swimming behaviour, of solution properties such as temperature~\citep{Balasubramanian2009}, contaminants~\citep{Zhao2013}, pH, and salt concentration are of critical importance to potential applications~\citep{Patra2013}. 

We consider a Janus polystyrene (insulating) spherical colloid of radius $a$, half coated by a Platinum  (conducting) shell. It is known that such colloids are active i.e. self-propel in hydrogen peroxide solution. This is due to gradients generated by the asymmetric decomposition of $\text{H}_2\text{O}_2$ on the Pt-coating and the interaction of the reactants and products with the sphere surface. 
%Our experience of phoretic mechanisms imply the self-propulsion is due to a slip layer of fluid near the sphere surface generated by these gradients.
\tl{In the rest of the paper we will call this process {\em self-phoresis}}.

Generically the catalytic decomposition of the hydrogen peroxide by the Platinum catalyst is given by 
\begin{equation}
\text{Pt} +  2{\text{H}_2\text{O}_2} \rightarrow \boxed{ \texttt{Intermediate-complexes} } \rightarrow \text{Pt} + 2\text{H}_2\text{O} + \text{O}_2 \ ,  \label{h2o2:decomposition:generic}
\end{equation}
however there is still some debate about the nature of the intermediate complexes~\cite{Hall1998,Hall1999,Hall1999a,Katsounaros2012}.
%could say this more accurately.

In this article, we outline a detailed calculation of the self-phoresis problem. Our approach is guided by 
%a comparison with 
the well studied problem of a phoretic motion of a colloid in an externally applied concentration gradient or electric field. 
To model the effect of the non-equlibrium chemical reaction sketched above on the motion of the Janus particle, we study the concentration fields of all the species involved in the reaction. The half coating of the colloid by catalyst is reflected by  inhomogeneous reactive boundary conditions on its surface.
% reflecting the half coating of the surface by catalyst. 
The reaction involves the production of charged intermediates which \tl{can also lead to changes in the electric potential on} the swimmer surface and hence the possibility of local electric fields. Our flexible calculation framework allows us to study a variety of 
different schemes for the reaction kinetics of the intermediate complexes. Using this we analyse in detail a scheme with both charged and uncharged pathways (see \textsf{Appendix  \ref{appA}}) whose results are consistent with all the behaviour observed in the recent experiments.
\section{The model}
\begin{figure}
\begin{center}
\subfigure{\includegraphics[scale=0.28]{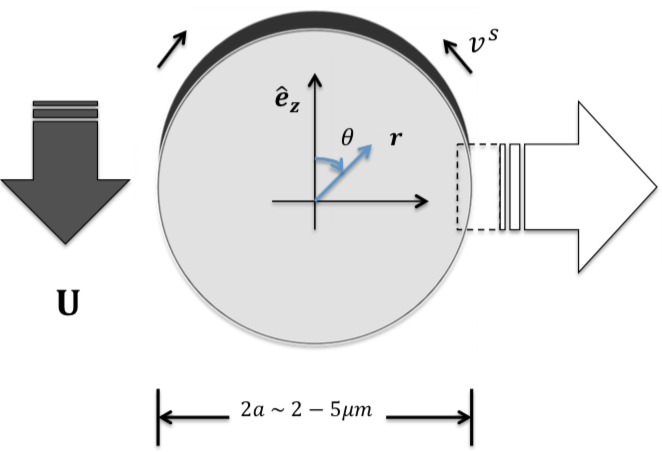}}
\subfigure{\includegraphics[scale=0.29]{matched_asymptotics3}}
\end{center}
\caption{Platinum-Polystyrene swimmer schematic with the domain decomposition of the phoretic problem.}
\label{fig:2}
\end{figure}

\tl{A Janus sphere of radius $a$ has the catalytic reaction of hydrogen peroxide decomposition 
occurring on its Pt coated half. 
We choose without loss of generality that the normal to the plane splitting the hemispheres is aligned with the $z$-axis [see Figure~\ref{fig:2}]. We propose a theoretical framework based on generally accepted properties of the reaction scheme for Pt catalysis of H$_2$O$_2$ degradation to water and O$_2$~\citep{Hall1998,Hall1999a,Hall1999}. A key feature of our analysis of self-propulsion is that it takes account of the existence of charged intermediates within the catalytic reaction scheme, namely protons and that the reaction rates varies with the Pt coating thickness  (see Figure \ref{fig:1})}.

\tl{The state of the system is therefore described by the local state of the Pt on the coated hemisphere, the electric potential, $\bar \Phi(\bar{\bs{r}})$, the fluid velocity, $\bar{\bfv}(\bar{\bs{r}})$, the local concentrations,  $\bar c_{hp} (\bar{{\bf r}}), \; \bar c_{o} (\bar{{\bs{r}}}), \; \bar c_{h} (\bar{{\bs{r}}})$ of H$_2$O$_2$, O$_2$ and H$^+$ respectively, i.e. the various reactive species, and the local concentrations, $\bar c_{oh}(\bar{\bs{r}})$, $\bar c_{s}(\bar{\bs{r} })$ of hydroxide and salt ions,  respectively}. The  background concentrations (far from the Janus sphere) of the salt, H$_2$O$_2$, H$^+$, and OH$^-$ are  \yhy{$c_s^\infty,\;  c_{hp}^\infty,\; c_h^\infty, \; c_{oh}^\infty$ respectively.}
\tl{Positions outside the Janus sphere (in the bulk) are represented by the vectors \yhy{$\bar{\bs{r}} = (\bar{x},\bar{y},\bar{z})$} (in Cartesian coordinates) while positions on the surface are parametrised by the unit vectors \yhy{$\nvec = \left( \sin\theta \cos \phi, \sin \theta \sin \phi, \cos \theta\right)$}.} We note that the vector \yhy{$\bar{\bs{r}}= ( \bar{r}, \theta , \phi)$ and $\nvec= \bs{\hat{e}}_{\bar{r}}$} in spherical polar coordinates. Each of the neutral species interacts non-electrostatically with the surface of the swimmer via a fixed {\em short ranged} potential energy \yhy{$\bar\Psi_n (\bar{\bs{r}})$}, that depends on the distance from the Janus sphere surface. \tl{The interaction range, \yhy{$L_{\text{eff}}$} is taken to be the same for all neutral species.}

\subsection{Equations of motion}
\tl{The relevant equations are Nernst-Planck equations~\citep{Probstein2003} for the  concentration of charged species, $\bar c_q$,
\beqa
\partial_t \bar c_q &=& - \bar{\nabla} \cdot \bar{\bfJ}_q \quad ; \quad \bar{{\bf J}}_q = - D_q \bar{\nabla} \bar c_q +  \bar{\bfu}_q \bar c_q \quad ; \quad \bar{\bfu}_q = \bar{\bfv}  - {D_q z_q e \over k_B T} \bar{\nabla} \bar \Phi \ ,
\eeqa
drift-diffusion equations ~\citep{Chandrasekhar1943} for the neutral species, $\bar c_n$, 
\beqa
\partial_t \bar c_n &=& - \bar{\nabla} \cdot \bar{\bfJ}_n \quad ; \quad \bar{{\bf J}}_n = - D_n \bar{\nabla} \bar c_n  + \bar{\bfu}_n \bar c_n \quad ; \quad \bar{\bfu}_n = \bar{\bfv} - {D_n \over k_B T} \bar{\nabla} \bar \Psi_n \ ,
\eeqa
Poisson's equation~\citep{J75} for the electric potential 
\beqa
\bar{\nabla}^2 \bar \Phi &=& - \sum_q { z_q e \bar c_q \over \epsilon}  \ ,
\eeqa
and the incompressible Navier-Stokes equations~\citep{Lamb} for the fluid velocity
\beqa
\rho \left(\partial_t \bar{\bfv}  \right. &+& \left. \bar{\bfv} \cdot \bar{\nabla} \bar{\bfv} \right)= \eta \bar{\nabla}^2 \bar{\bfv} - \bar{\nabla} \bar{p} + \bar{\bff} (\bar{\bs{r}}) \quad ; \quad \bar{\nabla} \cdot \bar{\bfv} = 0  \ ,  \\
\bar{\bff} (\bar{\bs{r}}) &=&  \sum_q {\Gamma}_q \bar c_q \left(\bar{\bfu}_q - \bar{\bfv} \right) + \sum_n  {\Gamma}_n \bar c_n \left( \bar{\bfu}_n - \bar{\bfv} \right)  \quad ; \quad \Gamma_q ={k_B T \over D_q} \; , \;  \Gamma_n = {k_B T \over D_n} \ , \nonumber 
\eeqa
%\beqa
%\partial_t \bar c_q &=& - \nabla \cdot \bfJ_q \quad ; \quad {\bf J}_q = - D_q \nabla \bar c_q +  \bfu_q \bar c_q \quad ; \quad \bfu_q = \bfV  - {D_q z_q e \over k_B T} \nabla \bar \Phi \\
%\partial_t \bar c_n &=& -\nabla \cdot \bfJ_n \quad ; \quad {\bf J}_n = - D_n \nabla \bar c_n  + \bfu_n \bar c_n \quad ; \quad \bfu_n = \bfV - {D_n \over k_B T} \nabla \bar \Psi_n \\
%\partial_t \bfV &+& \bfV \cdot \nabla \bfV = \eta \nabla^2 \bfV - \nabla P + \bff (\bs{r}) \quad ; \quad \nabla \cdot \bfV = 0  \\
%\bff (\bs{r}) &=&  \sum_q \Gamma_q \bar c_q \left(\bfu_q -\bfV \right)+ \sum_n  \Gamma_n \bar c_n \left( \bfu_n - \bfV \right)  \quad ; \quad \Gamma_q ={k_B T \over D_q} \; , \;  \Gamma_n = {k_B T \over D_n} \nonumber \\
%\nabla^2 \bar \Phi &=& - \sum_q { z_q e \bar c_q \over \epsilon}
%\eeqa
where , $\bar p(\bar{\bs{r}})$ is the hydrostatic pressure at $\bar{\bs{r}}$, $\eta$ is the viscosity, $k_B$ Boltzmann constant and $T$ temperature, $D_i$ is the diffusion coefficient of $i$'th solute and $z_i$ its valency if charged. These equations together with  the inhomogeneous boundary conditions (BC) on the surface of the Janus sphere and as $\bar r \rar \infty$ (see next section)  define a boundary value problem whose {approximate} solution is the subject of this paper.} \\

We consider the system in the steady-state (time derivatives equal to zero), the dynamics of the fluid around the swimmer in the zero Reynolds number (Re$=0$) limit of the Navier-Stokes equations for incompressible fluid flow. 
% and zero Pecl\'et number (Pe$=0$) for the solutes diffusion.
\tbl{In this paper we restrict ourselves to zero Pecl\'et number, equivalent to assuming that diffusion of the solutes occurs much faster than their convection by the flows generated by the Janus particle - very reasonable 
for the experimental systems we attempt to describe. }
\tl{Thus, the fluid velocity given by  $\bar{\bfv}  (\bar{\bs{r}}) = \bar{v}_r \bs{\hat{e}}_{\bar{r}} + \bar{v}_{\theta} \bs{\hat{e}}_{\theta}$ obeys the Stokes equation, while the solute concentration fields $\bar{c}_q, \ \bar{c}_n$ are governed by the steady-state drift-diffusion equations.}
\tl{\beqa
0 &=& \bar{\nabla} \cdot \bar{\bfv} \ ,  \\
\mathbf{0} &=&  \nabla \cdot \bs{\bar \Pi} + \bs{\bar f}= \eta \bar{\nabla}^2 \bar{\bfv} - \bar{\nabla} \bar{p} -  \sum_{i \in \text{ions}}e z_i \bar c_i \  \bar{\nabla} \bar \Phi - \sum_{j \in \text{non-ions}} \bar c _j \bar{\nabla} \bar \Psi_j  \label{eq:moment1}  \ , \\
0 &=& - \bar{\nabla} \cdot \bar{\bfJ}_q \quad ; \quad \bar{{\bf J}}_q = - D_q \bar{\nabla} \bar c_q - {D_q z_q e \bar c_q \over k_B T} \bar{\nabla} \bar \Phi \ ;  \qquad q \in \mbox{ions}  \ , \\
0 &=& - \bar{\nabla} \cdot \bar{\bfJ}_n \quad ; \quad \bar{{\bf J}}_n = - D_n \bar{\nabla} \bar c_n  - {D_n \bar c_n \over k_B T } \bar{\nabla} \bar \Psi_n \ ; \qquad n \in \mbox{non-ions}  \ .
\eeqa 
where we have defined $\bs{\bar \Pi}(\bs{\bar r})= \eta \left(\nabla \bs{v} +\nabla \bs{v}^T\right) - p \bs{\delta} $, the local hydrodynamic stress tensor.  }
%
%
%Before outlining the boundary conditions of these governing equations, in the next section, we will discuss the important notion of reaction rates variation with Pt metal thickness that sets the ionic fluxes on the Janus particle surface. 
%
%
%

%
\subsection{Boundary conditions}
\tl{The hydroxide and the salt ions are not involved directly in the catalytic decomposition of the fuel  (\ref{h2o2:decomposition:generic}) so we impose zero flux boundary conditions for their concentrations on the surface of the Janus particle, 
\begin{equation}
\bs{\hat{n}} \cdot \bar{\mathbf{J}}_{oh}|_{\bar{r}=a} = 0 =\bs{\hat{n}} \cdot \bar{\mathbf{J}}_{s,\pm}|_{\bar{r}=a} \quad . \label{OH:salt:flux:condition}
\end{equation} 
where the unit vector, $\nvec = \left( \sin\theta \cos \phi, \sin \theta \sin \phi, \cos \theta\right) = \bs{\hat{e}}_r$ in spherical polar coordinates. We define a catalyst coverage function, $K(\cos \theta)$ which is  $1$ on the Platinum hemisphere and zero on the polystyrene hemisphere,} 
\begin{equation}
K(\cos \theta ) = \left \{ \begin{array}{ll}
1,  & \ \  0 \leq \cos \theta \leq 1 \\
0,   &- 1 \leq \cos \theta < 0
\end{array} .
\right. \label{eq:kcos}
\end{equation} 

\tl{The presence of protons as intermediates of the fuel decomposition reaction  (\ref{h2o2:decomposition:generic}) and the  the variation of the reaction rates across the Pt-coated hemisphere leads to non-zero flux boundary conditions for the proton concentration on the surface of the Janus sphere
\begin{align}
\boldsymbol{\hat{n}} \cdot \bar{\mathbf{J}}_h|_{\bar{r}=a} &  = \bar{\mathcal{J}}_h(\theta) K(\cos\theta) \ , \label{H:flux:condition}
\end{align} 
where the proton current, $\bar{\mathcal{J}}_h$, varies with $\theta$ (position along the Pt-coated hemisphere). The specific form of the proton current $\bar{\mathcal{J}}_h$ will depend on the details of the reaction kinetics (see section \ref{sec:pt:thickness:rates} and Appendix \ref{appA}). 
%
%We defer the explicit expression of the proton flux to the electrophoretic and ionic diffusiophoretic section \ref{sec:ionic}. 
%
However, we note that $\bar{\mathcal{J}}_h > 0$ implies a chemical reaction producing protons while $\bar{\mathcal{J}}_h < 0$  implies a proton sink}. 

\tl{The fuel decomposition reaction involves the neutral species, H$_2$O$_2$ and O$_2$ giving rise to non-zero flux boundary conditions for their concentrations on the Janus-particle surface,
\begin{align}
\boldsymbol{\hat{n}} \cdot \bar{\mathbf{J}}_o|_{\bar{r}=a} & = \bar{\mathcal{J}}_{o} (\theta) K(\cos \theta)  \ , \\
\boldsymbol{\hat{n}} \cdot \bar{\mathbf{J}}_{hp}|_{\bar{r}=a} & = \bar{\mathcal{J}}_{hp} ( \theta ) K(\cos \theta)    \ ,
\end{align}
where $\bar{\mathcal{J}}_{hp}(\theta) < 0$ indicates H$_2$O$_2$ decomposition while $\bar{\mathcal{J}}_{o}(\theta) > 0$ indicates production of the O$_2$. Because of the variations in thickness of the Pt-coating, both $\bar{\mathcal{J}}_{hp}(\theta), \bar{\mathcal{J}}_{o}(\theta) $, defined in Appendix \ref{appA}, are functions of position along the Pt-coated hemisphere. 
%
%Expressions for these fluxes will be provided in the neutral self-diffusiophoretic section \ref{sec:inert}. 
}

\tl{All the concentrations, $\bar{c_i} (\bs{\bar r})$ decay to their background values, $\bar c_i^\infty$ as $\bar{r} \; \rar \;  \infty$.} 
%\\

\tl{We have Dirichlet boundary conditions for the electric potential on the particle surface
\begin{equation}
\bar{\Phi}(\bar{r} = a) \ = \  \bar{\varphi}_s (\theta) \ ,
\end{equation}
where $\bar{\varphi}_s$ is a possibly varying function over the swimmer surface. The potential, $\bar{\varphi}_s$ will in general be pH-dependent and will also depend on the 
particular reaction scheme of catalytic fuel decomposition.
For our analysis, it is sufficient to know the average value $\aver{\bar \varphi_s} = \frac{1}{2 \pi} \int d \cos \theta  \, \bar{\varphi}_s (\theta) $ and in the following 
we take $ \bar{\varphi}_s  \equiv \aver{\bar \varphi_s}$.
The potential $\bar{\varphi}_s$ can be related to the swimmer surface charge by double-layer models~\citep{Russel1992}. } \\

\par
The boundary conditions for the fluid velocity field are 
\yhy{\begin{equation}
\bar{\bfv}|_{\bar{r}=a} = \bar{\mathbf{U}} + \bar{\bs{\Omega}} \times \bar{\bs{r}}; \qquad \bar{\bfv}(\bar{r} \rightarrow \infty) = \mathbf{0} \quad ,
\end{equation} }
where \yhy{$\bar{\mathbf{U}},\bar{\bs{\Omega}}$} are respectively the total linear and angular propulsion velocities of the swimmer. 
\tl{These are unknown and their calculation is the goal of this paper}

\subsection{Constraints}

\emph{(Quasi-steady state condition)} As we study the system in a quasi-steady state, this requires that the average proton current on the swimmer surface vanishes,
\yhy{\begin{equation}
\oint_{\bar{r}=a} \bar{\mathcal{J}}_h (\theta) \ K(\cos \theta) \ \sin \theta \ d \theta  = 0   \   ,  \label{quasi:steady:constraint}
\end{equation} }
and note that this also guarantees conservation of the surface charge~\citep{Moran2011a}.  \\
%Note that the proton flux $\bar{\mathcal{J}}_h$ expression specified above in eqn. (\ref{proton:flux:dimens}) is obtained by imposing this constraint. Thereby automatically satisfying this requirement (see Appendix \ref{appA} for the details of its derivation). \\

\emph{(Swimming conditions)} We consider a freely swimming Janus particle with no external load on the colloid which requires that there is zero {\em total} force and torque on the swimmer:
\yhy{\begin{align}
\bar{\mathbf{F}} & = \oint_{\bar{r}=a} \bar{\bs{\Pi}} \cdot \bs{\hat{n}}  \ d \mathcal{S}_p + \int \bs{\bar f} d \mathcal{V}_p  = \mathbf{0} \ ,  \label{zero:force} \\
 \bar{\bs{T}} & = \oint_{\bar{r}=a} \bar{\bs{r}} \times \left( \bar{\bs{\Pi}} \cdot \bs{\hat{n}}\right) \ d \mathcal{S}_p + \int \bar{\bs{r}} \times \bs{\bar f} d \mathcal{V}_p   = \mathbf{0} \quad ,  \label{zero:torque}
\end{align}
where $d\mathcal{S}_p$ ($d\mathcal{V}_p$) is the differential surface (volume) element}. These two conditions uniquely determine both propulsion velocities $(\bar{\mathbf{U}},\bar{\bs{\Omega}})$\yhy{~\citep{Anderson1989}}. 

\tbl{The linearity of the Stokes equation and the limit of vanishing Pecl\'et number, mean that we can divide the linear and angular  velocities into non-electric, i.e. neutral diffusiophoretic, (due to the terms on the rhs of equation (\ref{eq:moment1}) depending on the $\bar \Psi_j$), and electric, i.e. ionic diffusiophoretic and electrophoretic, contributions (due to the terms on the rhs of equation (\ref{eq:moment1}) depending on $\bar\Phi$),  which can each be calculated separately,}
\yhy{\begin{align}
\bar{\mathbf{U} } & = \bar{\mathbf{U}}^e + \bar{\mathbf{U}}^d \ , \\
 \bar{\bs{\Omega}} & = \bar{\bs{\Omega}}^e + \bar{\bs{\Omega}}^d  \  ,
\end{align}  }
where \yhy{($\bar{\mathbf{U}}^e,\bar{\bs{\Omega}}^e$)} are electric and  \yhy{($\bar{\mathbf{U}}^d,\bar{\bs{\Omega}}^d$)} are non-electric. 
We expect (and indeed find) that the  neutral diffusiophoretic contribution to the propulsion is much smaller than the electrophoretic contribution. 
While we will later briefly outline the calculation of the neutral diffusiophoretic contribution to the propulsion velocity in section \ref{sec:inert}, in this article we will focus on the electrophoretic and ionic diffusiophoretic contributions. Detailed calculations of the neutral diffusiophoretic contribution can be found in the literature~\citep{AndersonPrieve1982,Golestanian2005,Golestanian2007,Michelin2014a}. 
% and hence drop the superscripts denoting the electrophoretic contibution ($\mathbf{U},\bs{\Omega}$). 

\tl{Due to the axisymmetry of the swimmer and the constraint of zero torque (\ref{zero:torque}), the angular velocity $(\bs{\bar \Omega})$ vanishes identically $(\bs{\bar \Omega}^e = \bs{0}, \ \bs{\bar \Omega}^d = \bs{0})$. Therefore in the following we will only consider the swimmer velocity $\bs{\bar U}$. }

%\subsection{Non-dimensionalization}
%
\subsection{Dependence of reaction rate constants on Pt-coating thickness\label{sec:pt:thickness:rates}} 
\tl{The cornerstone of our analysis in this paper is the identification of the fact that the reaction rate of H$_2$O$_2$ decomposition depends on the Pt-coating thickness~\citep{Ebbens2014}. A further observation is the well known presence of additional chemical pathways in the decomposition which involve charged intermediates, in particular protons,~\citep{Hall1998,Hall1999a,Hall1999}. 
%\citep{cite-pt-h2o2-papers}. 
These charged intermediates, in conjunction with the variation of Pt-coating thickness, allow an electric current to be established in the Pt shell due to varying decomposition rates of the hydrogen peroxide on different parts of the shell.} We approximate for simplicity that this thickness variation is linear in $\cos \theta$ , with a peak at the pole and the minimum at the equator, 
\begin{equation}
k_i(\theta) =  k^{(0)}_i + \sum_l k^{(l)}_i \ P_l \left(  \cos \theta \right) 
 \simeq  k^{(0)}_i + k^{(1)}_i \ \cos \theta  \ ,
\end{equation} 
\tl{where $k_i(\theta)$ is the reaction rate `costant' for $i$'th reaction step in reaction (\ref{h2o2:decomposition:generic}) above and $P_n(x)$ is the Legendre polynomial of order $n$. The Legendre moments $k_i^{(l)} =(l+1/2) \int_{-1}^1 k_i(\theta) P_l(\cos\theta) \, d\cos\theta$}. We assume weak variation ($k_i^{(1)} \ll k_i^{(0)}$) allowing us to work perturbatively in the variation. 
\tl{As long as there is a competition between a neutral pathway and a pathway involving charged intermediates, conservation of charge in the steady-state requires that the varying reaction rates across the Pt-coating lead to establishment of electric currents in the Pt shell.  This is described in detail for a particular reaction scheme involving protons in Appendix \ref{appA}, however the qualitative features of our results do not depend on the details of the scheme.} 

\section{Analysis}

%%%%%%%%%%%%%%%%%%%%%%%%%%%%%%%%

%%%%%%%%%%%%%%%%%%%%%%%%%%%%%%%%

Guided by current experiments, we analyse the coupled problem of the concentrations, electrostatic potential and fluid flow by  considering situations in which the length-scale of the interactions (Debye screening length, $\kappa^{-1}$ for charged species and \yhy{effective interaction range $L_{\text{eff}}$} for the neutral species) is small compared to the size (radius $=a$)  of the swimmer. We verify a posteriori that this is indeed the case. 
%Typically the interaction length-scale is $\sim 1\;$nm while the swimmer is of size $a \ge 100\; $nm. 
%\yhy{We consider the thin Debye-layer limit $\kappa a \gg 1$ where the interaction length-scale $\kappa^{-1} $ is much less than the swimmer size $a$. }
\yhy{The effective diffusiophoretic interaction range $L_{\text{eff}}$ for all the neutral solutes is defined $L_{\text{eff}}^2 = (\eta /k_BT)\bar{\mu}_d^{\ddagger}$, where $\bar{\mu}_d^{\ddagger} = {k_B T \over \eta} \int_0^\infty \rho \left( 1-e^{- \bar \Psi/ k_B T} \right) d \rho > 0$ is the characteristic diffusiophoretic mobility of the Janus particle.}
Hence the problem can naturally be viewed 
as one with two very separate length-scales with  small parameters $\lambda=1/(\kappa a),\yhy{\chi=L_{\text{eff}}/a}$ for charged and neutral species respectively. \tbl{A robust bound for comparison with experiment would be $\lambda \le 0.1$}.
\tl{A useful approach to multi-scale problems with a small parameter multiplying the differential operator of highest order, is the decomposition of the domain of the solution into a boundary layer, where the fields vary on the small $\mathcal{O} (\lambda)$ length-scale ($\mathcal{O}(\chi)$ for the diffusiophoretic contribution) and an outer domain where the characteristic length-scale is the size of the swimmer '$a$'. To do this most efficiently, we group the dimensionful quantities into useful dimensionless groups whose variation determines the behaviour of the system.}

\subsection{Self-electrophoresis and ionic self-diffusiophoresis\label{sec:ionic}}
In this section, we describe detailed calculations of the electrophoretic and ionic-diffusiophoretic contributions to the swimming velocity which is the main focus of the paper.

%\iffalse

\subsubsection{Dimensionless equations}
\tl{We non-dimensionalize the equations as follows.
The position vector \yhy{$\bar{\bs{r}}$} is measured in units of the swimmer size '$a$', concentrations $\bar c_i$ in units of the steady-state  background values $c_i^{\infty}$, 
% (we note that $\bar c_o$ is measured in units of $c_{hp}^\infty$)
electric potential $\bar \Phi$ in terms of the thermal voltage $(e\beta)^{-1}$ (with $\beta^{-1} = k_B T$, $k_B$ Boltzmann constant and $T$ temperature), ionic solute fluxes, $\mathbf{\bar J}_q$ in terms of {\color{black}$D_q \sum_i |z_i|^2 c_i^{\infty}/a$}, 
%neutral solute fluxes,$\mathbf{J}_n$ in units of  {\color{blue}$D_n c_{hp}^{\infty}/a$}, 
with $D_i$ the diffusion coefficient of $i$'th solute and $z_i$ its valency. The fluid flow velocity $\bs{\bar v}$ is rescaled by $\epsilon/(e^2\beta^2 \eta a)$, while the pressure $\bar p$  is rescaled by $\epsilon/(e^2 \beta^2 a^2)$. Hence we express dimensionless quantities (without overbar) in terms of the dimensionful (with overbar): $\bs{r} = (x,y,z)  = \bar{\bs{r}}/a, \
c_i  = \bar c_i  /c_i^\infty,  \
\Phi  = e \beta \bar \Phi, \  
\bs{v}  = \bar{\bf{v}} e^2 \beta^2 \eta a /\epsilon,  \ p = \bar{p} e^2 \beta^2 a /\epsilon$. }
%\beqa
%\bs{r} &=& (x,y,z)  = \bar{\bs{r}}/a \ ,\nonumber \\
%c_i  &=& \bar c_i  /c_i^\infty \ ,  \nonumber \\
%\Phi  &=& e \beta \bar \Phi \ , \nonumber \\
%%\Psi  &=& \beta \bar \Psi \nonumber \\
%%{\bf J}_i   &=& a  / D_i \sum_i |z_i|^2 c_i^{\infty}  \bar{\bf J}_i \nonumber \\
%\bs{v}  &=& \bar{\bf{v}} e^2 \beta^2 \eta a /\epsilon  \ , \nonumber \\
%p &=& \bar{p} e^2 \beta^2 a /\epsilon  \ . \nonumber 
%\eeqa

%\subsubsection{Solute number conservation and electric field}

It is useful for us to  define the dimensionless deviations of the solute concentrations, $C_{i}(\bs{r}) \equiv c_i (\bs{r}) - 1 =(\bar c_i/c^{\infty}_i) - 1  $ from their bulk values. 
%; $\bar c_i(\bar{\bs{r}})$ is the dimensional solute concentration. 
%In this paper we restrict ourselves to zero Pecl\'et number, equivalent to assuming that diffusion of the solutes occurs much faster than their convection by the flows generated by the Janus particle - very reasonable 
%for the experimental systems we attempt to describe.
%\par \noindent \textit{(a).} 
%The steady state equations for the concentration deviations of the charged species, 
Hence we obtain the following dimensionless equations of motion:
\begin{enumerate}[(1) ]
\item The steady-state equations for concentration differences of the charged species; protons $C_h$, hydroxide ions $C_{oh}$, and the salt $C_{s\pm}$,
\begin{equation} 
\nabla \cdot \mathbf{J}_i = 0; \quad  \mathbf{J}_{i}  = - \nabla C_{i}   -  z_i ( 1 + C_{i} )\nabla \Phi   \ ,
 \label{ions:nernstplanck}
\end{equation}
where $i \in \{h,oh,s\pm\}$.
%where $\nabla f \equiv (\partial f / \partial x, \partial f / \partial y, \partial f / \partial z)$, $C_{i} (\bs{r}) \equiv (c_i - 1) $ is the (dimensionless) $i$'th ionic solute concentration difference,  and $\Phi(\bs{r})$ is the electric potential. 
%
We consider only monovalent salts $|z_i|=1$.\\
\item \tl{The dimensionless Poisson's equation for the electric potential} $\Phi(\bs{r})$ , 
\begin{equation}
- \lambda^2 \ \nabla^2 \Phi  = \sum_{i \in \{ h,oh,s\pm \}}  \mathcal{Z}_i C_i  \  ,
\end{equation}
\item The dimensionless Stokes equations for the fluid velocity $\bs{v}  (\bs{r})$,
\beqa
0 &=& \nabla \cdot \bs{v} \ , \\
\mathbf{0} &=& \nabla \cdot \bs{\Sigma} = \nabla^2 \bs{v} - \nabla p -  \lambda^{-2} \sum_{i \in \text{ions}} \mathcal{Z}_i C_i \  \nabla \Phi   \ ,
\eeqa
\end{enumerate}
where the dimensionless parameters $\lambda$ and $\mathcal{Z}_i$ are defined as
\begin{equation}
\lambda^2 \equiv \left( \kappa a \right)^{-2}; \quad  \kappa^{-2} =  \frac{\epsilon \ k_B T }{ e^2 \sum_j |z_j|^2 c_j^{\infty} }=\frac{1 }{ 4 \pi l_B \sum_j |z_j|^2 c_j^{\infty} }; \quad  \mathcal{Z}_i = \frac{ z_i c_i^{\infty}}{\sum_j |z_j|^2 c_j^{\infty}}  \ , \label{Debye:alpha:definition}
\end{equation}
$\epsilon$ the permittivity of the solvent, and $e$ is the electronic charge. \tl{$\kappa^{-1}$ is the Debye screening length and  $l_B = e^2/4\pi \epsilon k_B T$ is the Bjerrum length~\citep{Russel1992}}. \tbl{We note that the stress $\bs{\Sigma}$ is the sum of the hydrodynamic stress tensor and the Maxwell stress tensor due to the interactions of the charged species with each other and the colloid surface.}

%
%\par \noindent \textbf{Swimming condition:} 
The zero total force condition which determines the propulsion velocity $\mathbf{U}^e$ becomes 
\begin{equation}
\mathbf{F} = \oint_{r=1} \bs{\Sigma} \cdot \bs{\hat{n}} \ \sin \theta \ d \theta  = \mathbf{ 0}  \ \, .   \label{force:free:ionic:contribution}
\end{equation}
%which determine uniquely the propulsion velocity $\mathbf{U}^e$. 

%\noindent We first perform a detailed calculation of the electric contribution to the swimming velocities before briefly outlining the (similar) calculation of the neutral diffusiophoeretic contribution. 

%%%%%%%%%%%%%%%%%%%%%%%%%%%%%%%%%
\subsubsection{Dimensionless boundary conditions}
\tl{For the  electric potential on the swimmer surface, 
%$\Phi(r=1) = \varphi_s(\theta) $ 
\begin{equation}
{\Phi}({r} = 1) \  = \  {\varphi}_s \ ,
\end{equation}
and  decays to zero in the bulk far from the swimmer, $\Phi(r\rightarrow\infty) = 0$ .} 
\par For the flow field  on the swimmer surface, \beq 
\bs{v}|_{r=1} = \mathbf{U}^e \ , 
\eeq and $\bs{v}(r\rightarrow \infty) = \mathbf{0}$ far in the bulk, where $\mathbf{U}^e$ is the electric contribution to the propulsion velocity.  
\par For the hydroxide and the salt concentrations, the zero flux boundary conditions  due to the impermeability of the Janus particle surface,
\begin{equation}
\bs{\hat{n}} \cdot \mathbf{J}_{oh}|_{r=1} = 0 =\bs{\hat{n}} \cdot \mathbf{J}_{s,\pm}|_{r=1} \quad . \label{OH:salt:flux:condition1}
\end{equation} 

For the proton concentration, the non-zero flux boundary condition,
\begin{equation}
\boldsymbol{\hat{n}} \cdot \mathbf{J}_h|_{r=1} = \mathcal{J}_h ( \theta ) K(\cos \theta)  \ ,  \label{H:flux:condition1}
\end{equation}
%where $K(\cos \theta)$ is the catalyst coverage function, $1$ on the Platinum hemisphere and zero on the polystyrene hemisphere. The unit vector, $\nvec = \left( \sin\theta \cos \phi, \sin \theta \sin \phi, \cos \theta\right)$ and vector $\bs{r}= (r,  \theta, \phi )$   and $\nvec= \bs{\hat{e}}_r$ in spherical polar coordinates. 
%(see e.g. the reaction kinetic model in \textsf{Appendix A})
\tbl{The essential mechanism which drives this process depends on the presence of a (1) varying proton flux (as a result of variation of  Pt thickness) which (2) averages to zero over the metallic hemisphere (due to charge conservation in the steady-state). In the limit of small linear variation in the thickness, this leads to  a proton flux of the general form
%\yhy{where the reaction kinetics specified proton flux (see Appendix \ref{appA}) reads
\begin{equation}
\mathcal{J}_h(\theta) = \gamma^{(1)} \left( 1 - 2 \cos \theta \right) K(\cos\theta) - \gamma^{(0)} \delta \left(\Phi + C_h \right) K(\cos \theta) \ . \label{proton:flux:definition} 
\end{equation}  
where both $\gamma^{(i)} \ne 0$. We note that $\gamma^{(1)}=0$ for a uniform thickness coating, and $\delta(\Phi+ C_h) = \left[ (\Phi + C_h) - \int_0^{\pi} ( \Phi  +  C_h ) \ K( \cos \theta) \ \sin \theta \ d  \theta \right]$ is the deviation of the {\em local} electric field and proton concentration from their surface average. $\gamma^{(0)}$ is a measure of the scale of typical production and consumption of protons across the metallic hemisphere. Since both terms on the rhs of eqn. (\ref{proton:flux:definition}) integrated over the surface give zero, the flux, $\mathcal{J}_h$ automatically satisfies the steady state requirement (\ref{quasi:steady:constraint}) and hence the conservation of total charge on the swimmer surface. }

\tl{Systems  which possess both properties above, with both $\gamma^{(i)} > 0$, will show all the qualitative behaviours described in this article, however their values will depend on the specific details of the chemical reaction scheme. 
A specific reaction scheme described in detail in Appendix \ref{appA} gives :
\begin{equation}
\gamma^{(0)} = \frac{  k_{\text{eff}}^{(h)} c_{hp}^{\infty} \ a }{ D_h \sum_i c_i^{\infty} }; \qquad  \gamma^{(1)} = \frac{\Delta k_{\text{eff}}^{(h)} c_{hp}^{\infty} \ a}{D_h \sum_i c_i^{\infty}}  \ . \label{gamma0:and:gamma1thickness}  
\end{equation}  
$k_{\text{eff}}^{(h)} >0 $ is the typical scale  of the average proton consumption and production while $\Delta k_{\text{eff}}^{(h)} >0$ is the scale of the \emph{difference} between the rates at the pole and equator (see Appendix \ref{appA} for their derivation from reaction kinetics) .
 }
 
\tl{We note  that the conservation of protons also requires  a relationship between the  pH of the solution and the potential on the surface of the Janus particle, which depends on the reaction kinetics (see Appendix \ref{appA});
\beq
\varphi_s = \varphi_s (c_h^\infty)\; ,
\eeq
leading to an estimate of the average swimmer surface charge $(\sigma_0(c_h^\infty))$ using the Gouy-Chapman model~\citep{Russel1992} of the interfacial double layer
\begin{equation}
\sigma_0 (c_h^\infty) = {e \kappa \over 2 \pi l_B} \sinh \left({\varphi_s \over 2}  \right) 
%\varphi_s = 2 \sinh^{-1} \left( \frac{2\pi l_B \sigma_0}{e \kappa} \right) 
\ , \label{zeta:gouy:chapman} 
\end{equation}  
where $l_B = e^2/4\pi \epsilon k_B T$ is the Bjerrum length, with $\epsilon$ the solution permitivity.}
%
%\yhyDelete{Equations (\ref{zeta:kinetics}) and (\ref{zeta:gouy:chapman}) uniquely determine the surface charge $\sigma_0$ given the solution pH and the catalytic reaction rate constants.} 
  \\
%
%%\yhyDelete{
%%\begin{equation}
%%\boldsymbol{\hat{n}} \cdot \mathbf{J}_h|_{r=1} = \mathcal{J}_h ( \theta ) K(\cos \theta)  \ ,  \label{H:flux:condition1}
%%\end{equation}
%%where $\mathcal{J}_h(\theta)$ is specified in eqn. (\ref{proton:flux:condition}) and $K(\cos \theta)$ is the catalyst coverage function, $1$ on the Platinum hemisphere and zero on the polystyrene hemisphere.  The unit vector, $\nvec = \left( \sin\theta \cos \phi, \sin \theta \sin \phi, \cos \theta\right)$ and vector $\bs{r}=r  \bs{\hat{e}}_r +  \theta \bs{\hat{e}}_\theta  + \phi \bs{\hat{e}}_\phi$   and $\nvec= \bs{\hat{e}}_r$ in spherical polar coordinates.  }
%%%\yhyDelete{The flux $\mathcal{J}_h ( \theta )$ will depend on details of the reaction kinetic model (see Appendix A). We note however that $\mathcal{J}_h > 0$ implies a chemical reaction producing protons while $\mathcal{J}_h < 0$ implies a proton sink. }
\iffalse
\begin{equation}
K(\cos \theta ) = \left \{ \begin{array}{ll}
1,  & \ \  0 \leq \cos \theta < 1 \\
0,   &- 1 \leq \cos \theta < 0
\end{array}
\right.
\end{equation}
\fi

%%%%%%%%%%%%%%%%%%%%%%%%%%%%%%%%

\begin{figure}
\begin{center}
\subfigure{\includegraphics[scale=0.28]{swimmer_sketch_asymptotic}}
\subfigure{\includegraphics[scale=0.29]{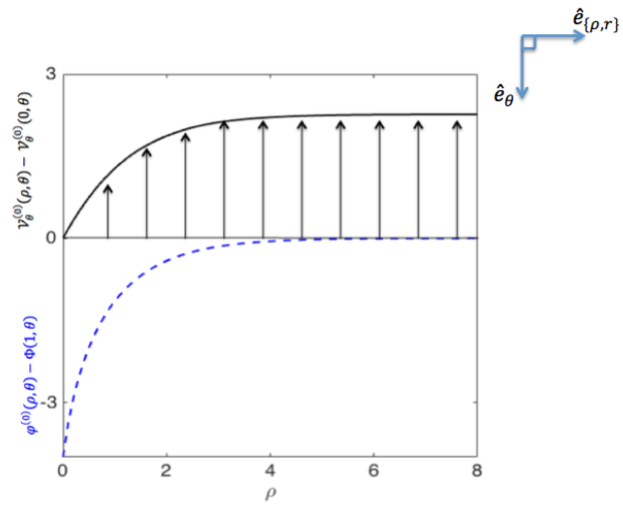}}
\end{center}
\caption{Profiles of flow and electric field within the (inner) Debye layer 
%(see equations (\ref{electro:pot:inner}, \ref{slip:flow}) 
with $\yhy{C_s^*}= \partial \yhy{C_s^*}/\partial \theta =  \partial \Phi/\partial \theta=1$ (see Appendix \ref{appB}). }
\label{fig:3}
\end{figure}

%%%%%%%%%%%%%%%%%%%%%%%%%%%%%%%%

\tl{In this electrostatic problem, the inner boundary-layer (double layer) fields, $H(\bs{r}) \in \{ c_i (\bs{ r}), \bs { v} (\bs{ r}),  {\bs \Phi} (\bs{ r})\}$ are expanded as}
\tl{
\begin{align}
H(r,\theta) & = \sum_n \lambda^n \ \mathcal{H}^{(n)} \left( \rho, \theta \right); \quad \rho = \frac{r - 1}{\lambda} \; ,
\end{align} 
while the outer fields, $H(\bs{r}) \in \{ c_i (\bs{ r}), \bs { v} (\bs{ r}),  {\bs \Phi} (\bs{ r})\}$ are expanded as
\begin{equation}
H(r,\theta) = H^{(0)}(r,\theta) + \sum_{n = 1}^{\infty} \lambda^n \ H^{(n)}(r,\theta) \ ,
\end{equation}  }
where $r$ is the bulk-scale coordinate. {\textcolor{black}{Similar expansions will apply for the self-diffusiophoretic problem, with $\lambda$ replaced by $\chi$.}}

\tl{The essence of the matched asymptotic method involves obtaining asymptotic expansions of the solutions of the equations in the limit $\lambda \rar 0$ for both the inner and outer fields and 
matching the results in the intermediate region:
\begin{equation}
\lim_{\lambda \rightarrow 0; \ \rho \rightarrow \infty} \{ \mathcal{H}^{(0)}_i \} \left( \rho, \theta \right) \ \ = \quad  \lim_{r \rightarrow 1; \lambda \rightarrow 0} \{  {H}_i^{(0)} \} (r,\theta) \quad  = \ \  \{ {H}_i \} (1,\theta)  \ .
\end{equation} 
In the next section, we will proceed to solve the outer problem in the limit of $\lambda = (\kappa a)^{-1} \rightarrow 0$, i.e thin double-layer limit where the swimmer radius $a$ is much larger than the Debye-layer thickness $\kappa^{-1}$. The details of the inner (Debye-layer) calculations~\citep{PrieveAnderson1984,Yariv2011} can be found in the Appendix \ref{appB} (see Figure \ref{fig:3}).}

\subsubsection{Outer concentration and electric  fields}
%\yhy{Having found the Debye-layer fields with all the necessary boundary conditions to describe the bulk fields, we are now set to solve for the bulk fields and hence finding the sought after ionic contribution of the swimmer propulsion velocity $\mathbf{U}^e$. }
%
 In the bulk, the fields vary over  length-scales comparable to the the swimmer size, with $\mathcal{O}(1)$ leading order fields and are expanded as
\yhy{\begin{equation}
H(r,\theta) = H^{(0)}(r,\theta) + \lambda \ H^{(1)}(r,\theta) + \cdots \ .
\end{equation} }
We drop the $^{(0)}$ superscript in the following as we will consider only the leading order terms $C^{(0)}, \Phi^{(0)}, v_i^{(0)},p^{(0)},$ in the expansions for the fields 
%(i.e. taking the Debye layer thickness as negligible compared to the swimmer radius).  \\
%\begin{figure}
%\begin{center}
%%\includegraphics[scale=.5]{proton_potential2} 
%\subfigure[]{\includegraphics[scale=.32]{Phi_N51020} }
%\subfigure[]{\includegraphics[scale=.24]{A1_radius} }
%\end{center}
%\caption{plots of ....}
%\end{figure}

%(\emph{Solute concentration and electric fields}) 
The leading order solute concentrations and electric potential outside the Debye-layer obey the equations
\begin{align}
\sum_{i \in \{ h,oh,s\pm \}} \mathcal{Z}_i C_i & = 0 \quad ,\label{outer:poisson:NP1}\\
\nabla \cdot \left[ \nabla C_i + z_i (1 + C_i ) \nabla \Phi \right] & = 0 \quad,  \label{outer:poisson:NP}
\end{align} 
where $\mathcal{Z}_i$ is defined in equation (\ref{Debye:alpha:definition}). \tl{It is useful for the rest of our analysis to treat all the ionic solutes together. Combining the two equations (\ref{outer:poisson:NP1},\ref{outer:poisson:NP}), 
%and the proton flux from the inner fields (\ref{proton:flux:outer:inner}), 
we obtain,} 
\yhy{\begin{align}
\nabla^2 C^* & = 0  \quad ,  \label{outer:conc:attenuation} \\
 \nabla \cdot \left( C^* \nabla \Phi \right) & = 0 \quad ,  \label{outer:pot:attenuation}
\end{align} 
where we have defined the sum of the deviations of concentration of all of the ionic solutes and its value at $r=1$.
\beqa
C^*(r,\theta) &=& 2 \sum_{i\in \{h,s+ \}} \mathcal{Z}_i \left( 1 + C_i(r,\theta)\right)\  ;  \quad \\ 
C_s^*(\theta) &\equiv &C^*(r=1,\theta)  \ .
\eeqa
\tl{The boundary conditions for  $\Phi$ and $C^*$ are obtained by matching to the inner solutions (see Appendix \ref{appB}), giving }
%The boundary conditions for  $\Phi$ and $C^*$ are  
\begin{align}
- \left. \hat{\bs{n}} \cdot \nabla C^* \right|_{r=1} & = \gamma^{(1)} \left( 1 - 2  \cos\theta \right) K(\cos\theta)  - \gamma^{(0)} \delta\left( \Phi + C_h\right) K(\cos\theta)  \ ,  \label{BC:ionic:solutes} \\
 - \hat{\bs{n}} \cdot \left( C^* \nabla \Phi \right|_{r=1} & = \gamma^{(1)} \left( 1 - 2  \cos\theta \right) K(\cos\theta)  - \gamma^{(0)} \delta\left( \Phi + C_h\right) K(\cos\theta)   \ ,   \label{BC:electric:potential}
\end{align} 
from eqns. (\ref{H:flux:condition1}) and (\ref{proton:flux:definition}). }\\ 
%
%

%(\emph{Flow field}) 
The fluid velocity field in the outer region obeys the equation
\begin{align}
& \nabla^2 \bs{v} - \nabla p + \nabla^2 \Phi \nabla \Phi = \mathbf{0} \ ,
\end{align}
with the slip boundary condition~\citep{PrieveAnderson1984}
\begin{align}
& \bs{v}(1,\theta)  = \mathbf{U}^e +  \left[ \zeta(\theta) \frac{\partial \Phi}{\partial \theta} + 4 \ln \cosh \left( \frac{ \zeta(\theta)}{4}\right)  \frac{\partial \ln \yhy{C_s^*}}{\partial \theta} \right] \bs{\hat{e}}_{\theta} \ ,
 \end{align}
and quiescent fluid far away from the swimmer, $\bs{v} \rightarrow \mathbf{0}$ as $r \rightarrow \infty$. \tl{The slip boundary condition for  $\bs{v}$ is obtained by matching to the inner solution (see Appendix \ref{appB}). }
%
%%%%%%%%%%%%%%%%%%%%%%%%
\begin{figure}
\begin{center}
\subfigure[]{\includegraphics[scale=.42]{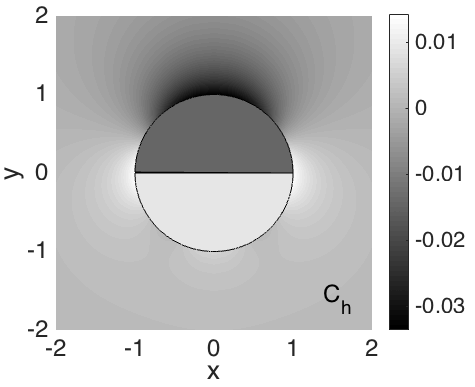} }
\subfigure[]{\includegraphics[scale=.4]{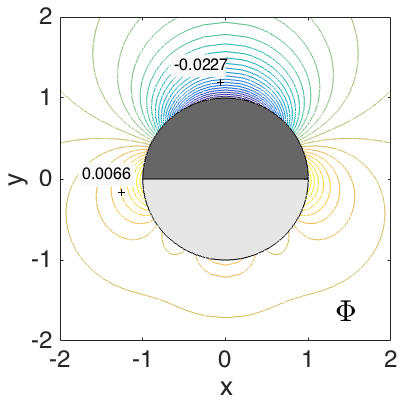} }
\end{center}
\caption{(a) \tl{Proton concentration deviation from the uniform background profile and (b) associated electric potential difference contours (both plots with $\mbox{pH} = 5.5$ for $10\% \ \mbox{H}_2\mbox{O}_2$ without salt and the system parameters in table (\ref{system:parameters})). The equipotential contours $0.0066$ and $-0.0227$ (in units of the thermal voltage $(k_BT/e) \approx 25$mVolts) are shown to indicate the electric pole-equator polarity. In both figures, the upper (dark) hemisphere is the Platinum cap. } }
\label{fig:4}
\end{figure}
%%%%%%%%%%%%%%%%%%%%%%%%
%
\subsubsection{Linear response and propulsion velocity} 

\tl{We note that with uniform coating, $k_i = k_i^{(0)}$, which implies $\gamma^{(1)} = 0$, the deviations of the electric potential and the ionic concentrations vanish ($\Phi = 0 = C_i$). The zeta potential for this trivial solution is 
\begin{equation}
\zeta_0 \ := \ \zeta \ = \ \varphi_s \label{zeta:pot:equilibrium}  \  . 
\end{equation}
In addition, this implies the fluid velocity field vanishes  $\bs{v} = \bs{0}$, and hence the contribution of self-electrophoresis to the propulsion velocity vanishes $\mathbf{U}^e = \bs{0}$. 
However, a varying thickness coating and  the consequent non-zero $\gamma^{(1)}$, lead to a qualitatively different scenario. To explore this we perform 
an expansion  to linear order in $\gamma^{(1)} / \gamma^{(0)}$ of the fields for the concentrations, fluid velocity, pressure and electric potential:  $\{C_i,\bs{v},p,\Phi \}$ for $\gamma^{(1)} \ll \gamma^{(0)}$,
%\begin{align}
%H  & = \gamma^{(1)} \ \gammapert{H} +  {\cal O}\left( (\gamma^{(1)})^2 \right) \ , 
%\end{align}  
where $\gamma^{(1)}, \gamma^{(0)}$ %, the scale of the variation in the proton flux variation 
are  defined in equation (\ref{gamma0:and:gamma1thickness}).}
\\
%

%\emph{(Solute concentration and electric fields)} 
\tl{We first expand the deviations of the concentrations and the electric field as
\begin{align}
H & = \gamma^{(1)} \ \gammapert{H} + {\cal O}\left( (\gamma^{(1)})^2 \right) \ , 
%C_i   & = \gamma^{(1)} \ \gammapert{C}_i + {\cal O}\left( (\gamma^{(1)})^2 \right) \ ,  \\
%C^*   & = \gamma^{(1)} \ \gammapert{C^{*}} + {\cal O}\left( (\gamma^{(1)})^2 \right) \ ,  \\
%\Phi  & = \gamma^{(1)} \ \gammapert{\Phi} + {\cal O}\left( (\gamma^{(1)})^2 \right) \ .
\end{align}
with $H \in \{ C_i, C^*, \Phi \}$ and keeping only linear terms. Substituting these perturbative fields into eqns. (\ref{outer:conc:attenuation},\ref{outer:pot:attenuation}), we find that at  leading order, $\gammapert{C^{*}}$ decouples from the electric potential field $\gammapert{\Phi}$ - with both obeying Laplace equations 
\begin{align} 
\nabla^2 \gammapert{C^{*}} &= 0  \ , \\
 \nabla^2 \gammapert{\Phi} \  &= 0 \ ,
\end{align} 
and the boundary conditions, from the matching with the inner solution, at this order are
\begin{align}
- \left. \hat{\bs{n}} \cdot \nabla \gammapert{C^{*}} \right|_{r=1}  = - \left. \hat{\bs{n}} \cdot \nabla \gammapert{\Phi} \right|_{r=1}  = \left( 1 - 2 x \right)K(x) - \gamma^{(0)} \delta(\gammapert{\Phi} + \gammapert{C_h}) K(x) \ ,  \label{BC:effective:C:Phi}
\end{align}
%\begin{align}
%- \left. \hat{\bs{n}} \cdot \nabla \tilde{C}^* \right|_{r=1} & = \left( 1 - 2 x \right) K(x) + \gamma^{(0)} \delta(\tilde{\Phi}) K(x),  \label{C_star_Bcondition} \\
% - \left. \hat{\bs{n}} \cdot \nabla \tilde{\Phi} \right|_{r=1} & = \left( 1 - 2 x \right) K(x) + \gamma^{(0)} \delta(\tilde{\Phi}) K(x),  \label{Phi_tilde_Bcondition}
%\end{align}
%
where $x = \cos \theta$, $\delta( \gammapert{\Phi}+ \gammapert{C_h}) =  \left[ (\gammapert{\Phi} + \gammapert{C_h} ) - \int_0^1 ( \gammapert{\Phi} + \gammapert{C_h} )dx \right]$.}

\tl{Now, the Laplace equations above for $\gammapert{C^{*}}, \gammapert{\Phi}$ in conjunction with the electroneutrality condition (\ref{outer:poisson:NP1}) imply (see Figure \ref{fig:4}) 
\begin{equation}
\gammapert{\Phi}(r,\theta) = \gammapert{C_h}(r,\theta) = \gammapert{C^{*}}(r,\theta) - 1= \sum_{l=0}^{\infty} A_l \ r^{-(l+1)} P_l(\cos \theta) \ ,  \label{Ch_Phi_solution}
\end{equation}  
where $P_l(\cos\theta)$'s are the Legendre polynomials. The unknown coefficients $A_l$'s are determined by the boundary conditions in equation (\ref{BC:effective:C:Phi}) above. }
%%%

Finally, the coefficients $A_l$ are obtained as a self-consistent system of equations,
\yhy{\begin{equation} 
\sum_{l=0}^{\infty} A_l (l+1) P_l \left( x \right) = \left( 1 - 2 x \right) K(x) -  2\gamma^{(0)} \sum_{l=0}^{\infty} A_l \left( P_l \left( x \right) - \int_0^1 P_l(x') dx' \right) \ K(x ) \ ,
\end{equation} }
where $x=\cos \theta$. 

\tl{Using the orthogonality of the Legendre polynomials, we obtain a linear system of equations for the coefficients $A_l$'s, 
%
%%%%%%%%%%%%%%%%%%%%%%%%%%
\begin{figure}
\begin{center}
\subfigure[]{\includegraphics[scale=.4]{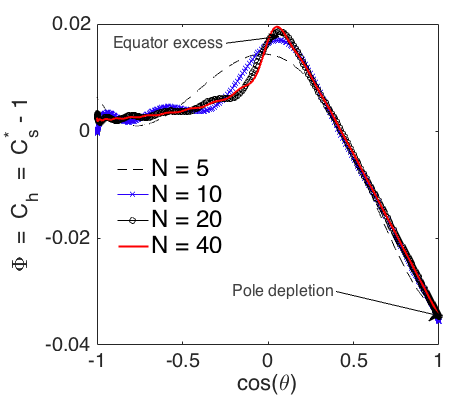} } 
%\subfigure[]{\includegraphics[scale=.24]{Phi_convergence.png} }
\subfigure[]{\includegraphics[scale=.4]{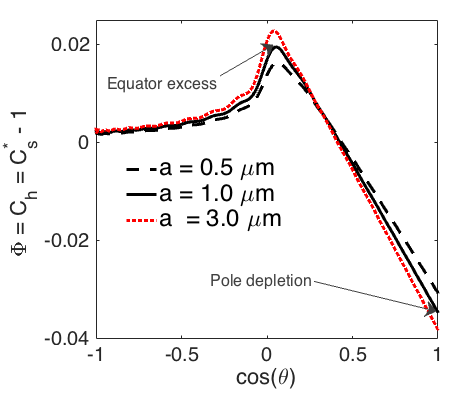} }
\end{center}
\caption{\tl{(a) The deviations of the surface electric potential, $\Phi(1,\theta)$, and ionic concentrations, $C_h(1,\theta), \ \yhy{C_s^*}(1,\theta) - 1$, from the uniform background values (for swimmer size $a = 1.00 \mu m$). We show the convergence of the solution as  the number  $N$ of the Legendre modes in eqn. (\ref{Ch_Phi_solution}) are increased, i.e ${\bf A} = \{A_0, \cdots , A_{N-1}, A_N \}$. (b) The deviations of the surface potential and ionic concentration as a function of swimmer size $a$ (truncating at $N=40$).} }
% Plot of the surface potential perturbation showing convergence as $N$ increases. } 
\label{fig:5}
\end{figure}
%%%%%%%%%%%%%%%%%%%%%%
%
\begin{equation}
\begin{pmatrix}
1  &    0  &    0   &  \cdots &     0  & \cdots\\
0  & M_{11}& M_{12}&  \cdots & M_{1l}& \cdots\\
0  & M_{21}& M_{22}&  \cdots & M_{2l}& \cdots\\
\vdots  & \vdots& \vdots& \ddots & \vdots& \hdots\\
0  & M_{n1}& M_{n2}&  \hdots & M_{nl}& \hdots\\
\vdots& \vdots& \vdots& \vdots& \vdots&  \ddots
\end{pmatrix}
\begin{pmatrix}
 A_0 \\ A_1\\ A_2\\ \vdots\\ A_l \\  \vdots
\end{pmatrix}
= \begin{pmatrix}
0 \\ \Lambda_1\\ \Lambda_2\\  \vdots \\ \Lambda_n \\ \vdots
\end{pmatrix}  \label{solution:Als}
\end{equation}
or more compactly 
\beq
\mathbb{M} \cdot {\bf A} = \bs{\Lambda} \; , 
\eeq
where ${\bf A} = (A_0, \ldots , A_{N}, \ldots)$, and  the matrix $\mathbb{M}$ and vector $\bs{\Lambda}$ entries are given by 
\yhy{\begin{align}
M_{nl} & = \delta_{nl} + \gamma^{(0)} \left( \frac{2n + 1}{n + 1 }\right) \int_0^1 P_n(x) \left(  P_l(x) - \int_0^1 P_l(x') dx' \right) \ dx \  , \\ 
\Lambda_n & = \frac{1}{2} \left( \frac{ 2n + 1 }{n + 1 }\right) \int_0^1 \left( 1 - 2 x \right) P_n(x) \ dx \  .
\end{align} }
}
The infinite linear system of equations (\ref{solution:Als}) above can be  solved approximately by truncating the infinite system after a finite number of components, reducing the description to the first  $N$ Legendre coefficients $A_l$'s. The approximate (numerical) solution requires inversion of  an  $N \times N$ matrix $\mathbb{M}$ (see Figure \ref{fig:5}). However, we can extract asymptotic regimes of this solution for $\gamma^{(0)} \ll 1$ and $\gamma^{(0)} \gg 1$. 
\tl{Note that $\gamma^{(1)} \ll \gamma^{(0)}$ in both limits.}\\
\par \noindent $\gamma^{(0)} \ll 1:$ In this regime, $A_l \sim \Lambda_l$ and 
\yhy{\begin{equation}
A_1 \sim - \frac{1}{8}
\end{equation}}
\par \noindent $\gamma^{(0)} \gg 1:$ In this regime, 
\begin{equation}
A_1 \sim  - \frac{\alpha}{\gamma^{(0)}}
\end{equation}
\tl{where $\alpha$ is a positive constant whose value can be determined numerically. The asymptotes show that the perturbations of $C_i$ and $\Phi$ decay to zero for large $\gamma^{(0)}$ (proportional to swimmer size) - when the diffusion time becomes large compared to the reaction time.}

\tl{In Fig.\ref{fig:5}(a), the deviations of the proton concentration $\gammapert{C_h}(1,\theta)$ and electric potential $\gammapert{\Phi}(1,\theta)$ on the surface from their bulk values are plotted showing the an excess at the equator  and depletion at the pole. Increasing the number of Legendre polynomial modes ($N$) improves the accuracy of the fields on the polystyrene hemisphere. The proton depletion (excess) at the pole (equator) is stronger for larger swimmer sizes (see Fig. \ref{fig:5}(b)).  }
\begin{table*}
\yhy{
\centering
\begin{tabular}{l| c | c | c}
\hline 
Description & Symbol & Value & Units (SI) \\
\hline \hline
\tbl{Boltzmann energy scale (at $300$  K)} & $k_B T$ & $4.05 \times 10^{-21}$  & $\mbox{J}^{}$  \\
%(room) Temperature & $T$  & $300$ & K \\
Permittivity (water) & $\epsilon$ & $6.90 \times 10^{-10}$ & $\mbox{C}\mbox{V}^{-1}\mbox{m}^{-1}$ \\
 Electronic charge & $e$ & $1.60 \times 10^{-19}$ & C  \\
 Average surface charge density (at zero salt conc.) & $\sigma_0$  &  $1.60 \times 10^{-3} $ & Cm$^{-2}$  \\
 Viscosity of water (at 300K)  & $\eta$ & $8.9 \times 10^{-4}$ & Nm$^{-2}$s$^{-1}$  \\
 Diffusiophoretic characteristic mobility & $\bar{\mu}_d^{\ddagger}$ & $4.57 \times 10^{-38}$ & m$^5$ s$^{-1}$  \\
  Peroxide $[\mbox{H}_2\mbox{O}_2]$ diffusion coefficient & $D_{\text{hp}}$ & $ 6.60 \times 10^{-10} $  & $\mbox{m}^2 \mbox{s}^{-1}$ \\
 Oxygen $[\mbox{O}_2]$ diffusion coefficient &  $D_{\text{o}}$ & $ 2.00 \times 10^{-9}$ & $\mbox{m}^2 \mbox{s}^{-1}$\\
 Protons $[\mbox{H}^+]$ diffusion coefficient &  $D_{\text{h}}$ & $9.30 \times 10^{-9}$  & $\mbox{m}^2 \mbox{s}^{-1}$\\
 % Hydroxide $[\mbox{OH}^-]$ diffusion coefficient &  $D_{\text{oh}}$ & $5.3 \times 10^{-9}$ & $\mbox{m}^2 \mbox{s}^{-1}$\\
Swimmer radius & $a$ & $1.00 \times 10^{-6}$ & m \\
%Neutral solutes interaction range & $L_{\mbox{eff}}$ & $1.00 \times 10^{-10} $ & m  \\
H$_2$O$_2$ decomposition reaction rate $(\mathcal{K} := k_{\text{eff}}^{(hp)}c_{hp}^{\infty})$   & $\mathcal{K}$  & $3.00 \times 10^{22}$ & $\mbox{m}^{-2} \mbox{s}^{-1}$ \\ 
\citep{Ebbens2014,Brown2014} & & & \\ 
$10$\% w/v H$_2$O$_2$ number concentration & $c_{hp}^{\infty}$ & $1.76 \times 10^{27} $ &  m$^{-3}$  \\
%H$_2$O$_2$ decomposition rate $(-k_{\text{eff}}^{(hp)}c_{hp}^{\infty})$  coefficient & $k_{\text{eff}}^{(hp)}$ & $1.70 \times 10^{-5} $ & m s$^{-1}$  \\ 
Effective proton absorption/release rate  ($\sim 0.3 \% \  {\mathcal{K}}$)  & $k_{\text{eff}}^{(h)} c_{hp}^{\infty}$ & $1.00 \times 10^{20}$ & $\mbox{m}^{-2} \mbox{s}^{-1}$ \\
Proton pole-to-equator rate `\textit{difference}' ($\sim 0.09 \% \  {\mathcal{K}}$) & $\Delta k_{\text{eff}}^{(h)} c_{hp}^{\infty}$ & $2.70 \times 10^{19}$ & $\mbox{m}^{-2} \mbox{s}^{-1}$ \\ 
\hline 
\hline
\end{tabular} 
\caption{System parameters} 
 \label{system:parameters}
 }
 \end{table*} \\
%

%(\emph{Flow field})
 \tl{The calculated coefficients $A_l$'s above determine the slip velocity and we can now solve the Stokes flow problem. Hence, as above,  we expand the velocity and pressure fields about the trivial solution $\bs{v} = \mathbf{0}, \ p = p_{\infty}$ ,
\begin{align}
\bs{v} &= \gamma^{(1)} \gammapert{\bs{v}} + \cdots; \\
 p - p_{\infty} &= \gamma^{(1)} \gammapert{p} + \cdots
\end{align} 
and the propulsion velocity about the stationary colloid, $\mathbf{U}^e = \mathbf{0}$
\begin{align}
\mathbf{U}^e &= \gamma^{(1)} {\gammapert{\mathbf{U}^e}} + \cdots 
\end{align} 
Then, the Stokes equations become 
\begin{align}
& \nabla^2 \gammapert{\bs{v}} - \nabla \gammapert{p}   = \mathbf{0} \ ; \qquad \nabla \cdot \gammapert{\bs{v}} = 0  \ ,
\label{eq:stokes_hom}\end{align}
with the slip boundary condition from matching to the inner solution,
\begin{equation}
\gammapert{\bs{v}}(1,\theta) = \gammapert{\mathbf{U}^e}  + \mu_e \frac{\partial \gammapert{\Phi}}{\partial \theta} \bs{\hat{e}}_{\theta}   \ ,
\end{equation} 
where $\mu_e =  \zeta_0  + 4 \ln  \cosh \left( \zeta_0 /4 \right)  $. Recall that $\zeta_0 = \varphi_s$ is the zeta potential for the trivial solution with $\gamma^{(1)}=0$.}   \\

\begin{figure}
\begin{center}
\subfigure[]{\includegraphics[scale=.2]{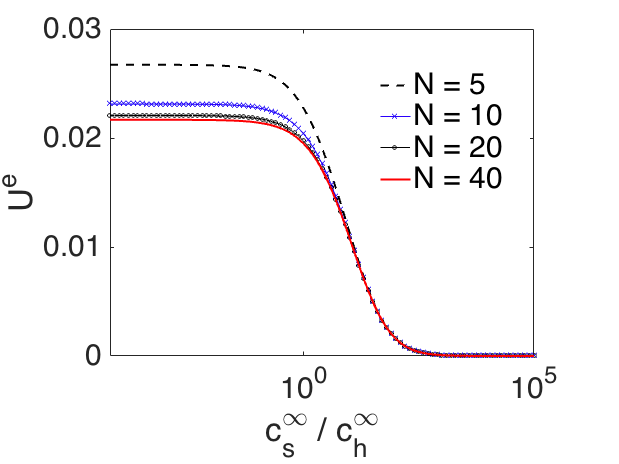}}
\subfigure[]{\includegraphics[scale=.2]{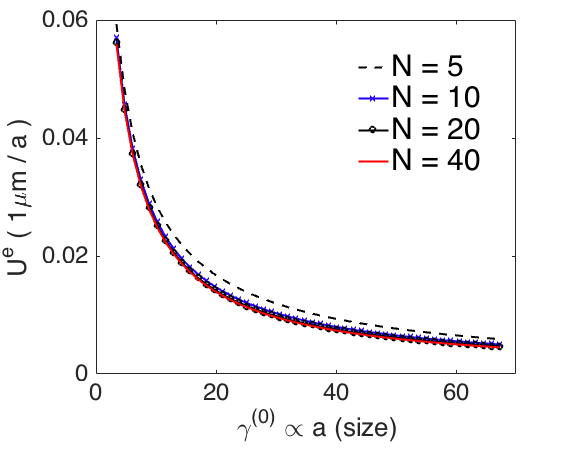}}
\subfigure[]{\includegraphics[scale=.24]{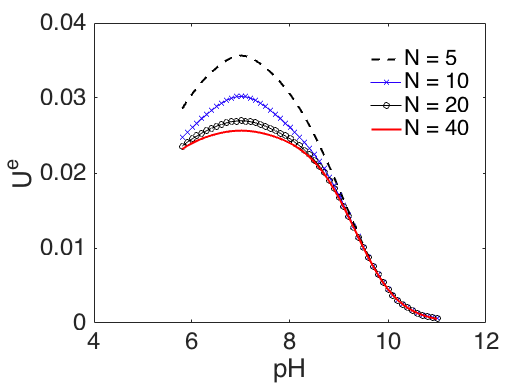}}
\end{center}
\caption{(Electrophoretic and ionic-diffusiophoretic contribution): (a) The propulsion speed $U^e$ as a function of the salt concentration $C_s^{\infty}$. (b) The propulsion speed $U^e$ decay with increasing size $a$. (c) The speed $U^e$ against the solution pH (\textcolor{black}{we expect that the charge balance maybe more complicated and the reaction kinetics are known to change with the solution pH}~\citep{liu2014ph,mckee1969catalytic}).}
\label{fig:7}
\end{figure}

\tl{Solving the homogeneous Stokes equations (\ref{eq:stokes_hom}) with these boundary conditions gives the following structure for the flow generated by the electrophoretic and ionic diffusiophoretic contributions:
\begin{align}
\bs{v}(\bs{r}) & = \gamma^{(1)}\gammapert{\bs{v}} =  B_2 \ \Big [ - \partial_{z}  \bs{G} (\bs{r}) \Big ] +  B_1 \ \bs{D}(\bs{r})   +  B_3 \ \left[ \partial_z^2 \bs{G} (\bs{r})\right]  + \ \mathcal{O}(r^{-4}), \label{flow:free:space:solution}
\end{align} 
expressed in terms of the leading order fundamental singularities of the Stokes equation:
\begin{equation}
\bs{G}(\bs{r}) = \frac{\hat{\bs{e}}_z}{r} + \frac{\bs{r}\bs{r} \cdot \hat{\bs{e}}_z}{r^3} \ ; \qquad \bs{D}(\bs{r}) = 3 \frac{\bs{r}\bs{r} \cdot \hat{\bs{e}}_z}{r^5} - \frac{\hat{\bs{e}}_z}{r^3} \ , \label{eq:stokes_sing}
\end{equation}
with the strengths given by
\begin{align}
B_1 = - \frac{1}{3} \mu_e \gamma^{(1)} A_1 \left( 1 - \frac{3}{2}\frac{A_3}{A_1} \right) ; \quad B_2 = \frac{3}{2} \mu_e  \gamma^{(1)} A_2; \quad B_3 = \frac{5}{4} \mu_e  \gamma^{(1)} A_3  \ ,
\end{align} 
where the $A_l$ are obtained from solving equations (\ref{solution:Als}).
% We define the leading (zeroth) order mobility and the swimmer surface average zeta potential 
%\begin{equation}
%\mu_e  = \zeta_0  + 4 \ln \cos\left( \frac{ \zeta_0}{4}\right) \ \ .
%\end{equation}
%
%
%It therefore follows that, i
Imposing the constraint of zero total force, equation (\ref{force:free:ionic:contribution}), leads to an expression for the 
electrophoretic and ionic-diffusiophoretic contributions to the propulsion velocity,
\begin{equation}
\mathbf{U}^e = - \frac{2}{3} \mu_e \gamma^{(1)}  A_1  \ \bs{\hat{e}}_z  \ . \label{electro:contr}
\end{equation}  
%with a correction of $\mathcal{O}\left( \Delta \zeta_0/\left< \zeta_0\right> \right)$. 
Written in dimensional form (see Figure \ref{fig:7}), %the linear contribution reads
\begin{equation} 
\bar{\mathbf{U}}^e = - \frac{1}{3} \left( \frac{k_BT}{e} \right)^2 \frac{\epsilon}{\eta} \frac{A_1 \ \Delta k_{\text{eff}}^{(h)} \ c_{hp}^{\infty} }{D_h \left( c_h^{\infty} + c_s^{\infty} \right)}\left[ \frac{e  \bar{\zeta}_0  }{k_BT} + 4 \ln \cosh\left( \frac{e  \bar{\zeta}_0 }{4 k_BT}\right) \right] \bs{\hat{e}}_z \ , \label{propulsion:velocity}
\end{equation} 
where $ \bar{\zeta}_0  = (k_BT/e) \zeta_0 $ is the average zeta swimmer average zeta potential. 
A plot of this electrophoretic contribution against the solution salt concentration (ionic strength) is shown in Fig. \ref{fig:7}(a). As expected, this contribution is strongly sensitive to salt concentration. Interestingly, the swimmer speed is only weakly dependent on pH  (see Fig. \ref{fig:7}(c)) under weakly acidic conditions (high $c_h^\infty$). This is due to the competition between the dependence on $c_h^\infty$ of $A_1$ (decreases with $c_h$), $\mu_e$ (increases with $c_h$) and the denominator of the expression for $\mathbf{U}^e$ (increases with $c_h$). This is consistent with recent experiments~\citep{Brown2014} which showed a minor reduction of swimming speed on addition of sodium hydroxide (NaOH). Furthermore, the propulsion speed is inversely dependent on swimmer size, $a$ for  large swimmer sizes as shown in Fig. \ref{fig:7}(b). This is consistent with the experimental observation of $\sim 1/a$ propulsion velocity decay for large swimmer sizes~\citep{Howse2007}. }
%Though, the origin of the inverse decay here is that the proton diffusion time becomes comparable to the reaction time for larger swimmer size and hence the proton exchange (oxidation/reduction) becomes effectively local on the Pt cap.}
%
\iffalse
\begin{figure}
\begin{center}
\includegraphics[scale=.25]{flow_str4}
\caption{The far-field flow streamlines (retaining $A_1$ and $A_2$ only) - a \textit{pusher} force-dipole.}
\end{center}
\label{fig:6}
\end{figure}
\fi

\subsection{Self-diffusiophoresis\label{sec:inert}}

\tl{In this section, we outline a solution of the equations of motion for the neutral solutes in the outer region $(\chi = L_{\text{eff}}/a \rightarrow 0)$ to calculate the neutral diffusiophoretic contribution to the propulsion velocity, $\mathbf{U}^d$. Detailed calculations for the inner interaction layer where the fields varies at the lengthscale $L_{\text{eff}}$ can be found in the literature~\citep{AndersonPrieve1982,Golestanian2005,Golestanian2007,Howse2007,Michelin2014a}. 
}

\tl{Here since there is a finite propulsion velocity, $\mathbf{\bar U}^d \ne 0 $, for the uniformly coated system, $k_i^{(0)}\ne 0, k_i^{(1)}=0$, then a weak variation of rates due to a varying thickness$,k_i^{(0)}\gg k_i^{(1)}$ leads to a small correction which we can ignore. Hence we set $ k_i^{(1)}=0$ for the rest of this section.}

\subsubsection{Dimensionless equations}
%
%We non-dimensionalize the equations as follows.
\tl{The position vector $\bar{\bs{r}}$ is measured in units of the swimmer size '$a$', concentrations  $\bar c_i$ in units of the steady-state  background values $c_i^{\infty}$ (note that $\bar c_o$ is measured in units of $c_{hp}^\infty$), the short-ranged interaction potential of solutes with the Janus sphere, $\bar \Psi$ in terms of the thermal energy scale $\beta^{-1} = k_B T$, 
%ionic solute fluxes, $\mathbf{J}_q$ in terms of {\color{blue}$D_q \sum_i |z_i|^2 c_i^{\infty}/a$}, 
neutral solute fluxes, $\bar{\mathbf{J}}_n$ in units of  {\color{black}$D_n c_{hp}^{\infty}/a$}, with $D_i$ the diffusion coefficient of $i$'th solute.
% and $z_i$ its valency. 
 The fluid flow velocity $\bar{\bf{v}}$ is rescaled by $\bar{\mu}_d^{\ddagger} c_{hp}^\infty/a$, where $\bar{\mu}_d^{\ddagger}$ is the characteristic diffusiophoretic mobility}, the pressure $\bar{p}$  is rescaled by \tl{$\bar{\mu}_d^{\ddagger} \eta c_{hp}^\infty/ a^2$. Hence, the dimensionless quantities (no overbar) are expressed in terms of dimensional ones (with overbar) as follows $\bs{r} = (x,y,z)  = \bar{\bs{r}}/a, \  c_i  = \bar c_i  /c_i^\infty,  \  \Psi  = \beta \bar \Psi,  \  \bs{v}  = \bar{\bf{v}} a /c_{hp}^\infty \bar{\mu}_d^{\ddagger}, \  p = \bar{p} \ a^2 / c_{hp}^\infty \bar{\mu}_d^{\ddagger} \eta$. }
As before we define the dimensionless difference of the concentrations from their bulk values as $C_{i}(\bs{r}) \equiv c_i (\bs{r}) - 1 =(\bar c_i/c^{\infty}_i) - 1 $.
%; $\bar c_i(\bs{r})$ is the dimensional solute concentration.
% and we restrict ourselves to zero Pecl\'et number, equivalent to assuming that diffusion of the solutes occurs much faster than their convection by the flows generated by the Janus particle.
%The molecules of the neutral species (hydrogen peroxide and oxygen) diffuse freely within the bulk, %and interact non-electrostatically with the surface of the swimmer via a potential energy,  that depends on the distance from the janus sphere surface.  We denote the potential for the $i$th neutral specie by $\Psi_i$. 
%and therefore their concentration fields obey the Laplace equations
% in a potential neutral hydrogen peroxide and oxygen diffuse while interacting non-electrostatically with the swimmer surface 
%The neutral solutes diffuse freely outside the interaction region ($r > L/a$)

The dimensionless equations for $r>1$ in the outer region are thus Laplace equations for the concentration deviations
\begin{align}
\nabla^2 C_o &= 0 \ ,  \label{o2:laplace} \\
 \nabla^2 C_{hp} &= 0  \ ,   \label{hp:laplace}
\end{align} 
%\begin{equation}
%\nabla^2 C_i = 0 \ ;  \qquad  i \in \{ o, hp \} \ .
%\end{equation}
%\begin{equation} 
%\nabla \cdot \mathbf{J}_i = 0; \quad \mathbf{J}_{i}  = -  \nabla C_{i}   -    ( 1 +  C_{i}) \nabla \Psi_i    
% \label{neutral:solute:conservation}
%\end{equation}
and the Stokes equations for the fluid velocity , $\bs{v}  (\bs{r}) $
\beqa
\mathbf{0} &=& \nabla \cdot \bs{\Pi} = \nabla^2 \bs{v} - \nabla p \ ; \qquad   0   =  \nabla \cdot \bs{v}   \ ,
\eeqa
%\beqa
%\mathbf{0} &=& \nabla \cdot \bs{\Pi} = \nabla^2 \bs{v} - \nabla p  - \chi^{-2} \sum_{j \in \text{non-ions}} C_j \nabla \Psi_j \\
%0 &=& \nabla \cdot \bs{v} 
%\eeqa
where $p(\bs{r})$ is the hydrostatic pressure at $\bs{r}$~\citep{Anderson1989,Golestanian2005,Golestanian2007,Howse2007,Michelin2014a} . 

%The dimensionless parameter $\chi^{-2} =  \left(a /L \right)^2 $  where $L$ is the range of the non-electrostatic interaction of the neutral solutes ($[\mbox{H}_2\mbox{O}_2]$ and $[\mbox{O}_2]$) with the swimmer surface.  

%where $\Psi_i$ is the mean interaction energy of a molecule of  the $i$'th solute  with the swimmer surface. 
%Their concentration attenuation fields $\{ C_{hp}, C_o \}$ 

\subsubsection{Dimensionless boundary conditions}

Matching with the inner layer~\citep{Anderson1989,Golestanian2005,Golestanian2007,Howse2007,Michelin2014a}, gives rise to non-zero flux  boundary conditions for hydrogen peroxide and oxygen
\begin{align}
 - \left. \partial_r C_o \right|_{r=1} & = \mathcal{J}_{o}(\theta) K(\cos \theta) \ = \quad \frac{D_{hp}}{2D_o}   \mathcal{K}_{\text{eff}}^{(hp)} \Big(1 + C_{hp} \Big) K(\cos \theta)  , \ \\
- \left. \partial_r C_{hp} \right|_{r=1} & =  \mathcal{J}_{hp}(\theta) K(\cos \theta) = - \mathcal{K}_{\text{eff}}^{(hp)} \Big( 1 + C_{hp} \Big) K(\cos \theta)  \ , 
\end{align}
%\begin{align}
%\boldsymbol{\hat{n}} \cdot \mathbf{J}_o|_{r=1}  & = \mathcal{J}_{o} (\theta) K(\cos \theta) \ , \\
%\boldsymbol{\hat{n}} \cdot \mathbf{J}_{hp}|_{r=1}  & = \mathcal{J}_{hp} ( \theta ) K(\cos \theta) \ ,
%\end{align}
and vanishing concentration deviations  far from the swimmer $C_o, \ C_{hp}\rightarrow 0$ as $ r \rightarrow \infty$. $K(\cos \theta)$,  the catalyst coverage function, is  $1$ on the Platinum hemisphere and zero on the polystyrene hemisphere.
%
%\yhy{In this section, we outline the dimensionless boundary conditions for the neutral solutes diffusion and Stokes equations. 
\tl{From the reaction kinetics in Appendix \ref{appA}, we obtain non-zero fluxes for hydrogen peroxide and oxygen
\begin{align}
\mathcal{J}_o(\theta) & = \frac{1}{2} \left(\frac{k_{\text{eff}}^{(hp)} \bar{c}_{hp} a}{D_o c_{hp}^{\infty}} \right) K(\cos\theta) \ = \quad \frac{D_{hp}}{2D_o}   \mathcal{K}_{\text{eff}}^{(hp)} \Big( 1 + C_{hp}(1,\theta) \Big) K(\cos \theta)  \ , \\ 
\mathcal{J}_{hp}(\theta) & = -\left( \frac{k_{\text{eff}}^{(hp)} \bar{c}_{hp} a}{D_{hp} c_{hp}^{\infty}} \right) K(\cos\theta) =   - \ \mathcal{K}_{\text{eff}}^{(hp)} \Big( 1 + C_{hp}(1,\theta) \Big) K(\cos \theta)  \ .
\end{align}
We have defined dimensionless $\mathcal{K}_{\text{eff}}^{(hp)} =  k_{\text{eff}}^{(hp)} a / D_{hp}$, where $k_{\text{eff}}^{(hp)}\bar{c}_{hp}>0$ is the effective rate of consumption of the hydrogen peroxide (see Appendix \ref{appA} for details of the derivation).} %\subsubsection{Momentum conservation}
%We consider the dynamics of the fluid around the swimmer in the zero Reynolds number (Re$=0$) limit of the Navier-Stokes equations for incompressible fluid flow. 

The boundary conditions for the fluid velocity are 
\tl{\begin{equation}
\bs{v}|_{r=1} = \mathbf{U}^d   + \bs{v}^d_{\text{slip}} \  ; \qquad \bs{v}(r\rightarrow \infty) = \mathbf{0} \quad ,
\end{equation}
where $\mathbf{U}^d$ is the neutral self-diffusiophoretic contribution to the propulsion velocity. $\bs{v}^d_{\text{slip}} = \sum_{i\in \{o,hp\}} \mu_d^{(i)} \left( \mathds{1} - \hat{\bs{n}} \hat{\bs{n}} \right) \cdot \nabla C_{i}$ is the self-diffusiophoretic slip velocity obtained by matching with the inner solution~\citep{AndersonPrieve1982,Anderson1989} and $\mathds{1}$ is a unit matrix. 
The dimensionless self-diffusiophoretic mobility is given by $\mu_d^{(i)} =  \lim_{ \rho' \rightarrow \infty} \int^{\rho'}_0 \rho \left[ 1 - e^{-\Psi_{i}(\rho) }  \right] d\rho$~\citep{Anderson1989}.
% with dimensional characteristic value of $\mu_d^{\ddagger}$. 
}

\tl{Finally, the zero total force condition on the swimmer
\begin{equation}
\mathbf{F} = \oint_{r=1} \bs{\Pi} \cdot \bs{\hat{n}} \ \sin \theta \ d \theta  = \mathbf{ 0} \ ,
\end{equation}
determines  the diffusiophoretic propulsion velocity $\mathbf{U}^d$.}

\subsubsection{Outer concentration fields}
\tl{The general solution of the Laplace equations (\ref{o2:laplace},\ref{hp:laplace}) for the neutral solutes is of the form
\begin{align}
C_o (r,\theta)  & = \sum_{l=0}^{\infty} \left( \frac{ D_{hp}}{2D_o}\right) W_l \ \frac{P_l(\cos \theta)}{r^{l+1}}  \ , \\
C_{hp}(r,\theta)  & = - \sum_{l=0}^{\infty} W_l \ \frac{P_l(\cos \theta)}{r^{l+1}} \ , 
\end{align} 
where $P_l(\cos\theta)$ are the Legendre polynomials and note that we have used the fact that $D_{hp}\mathcal{J}_{hp} + 2 D_o \mathcal{J}_o = 0$.  The amplitudes $W_l$'s, are determined from either of the boundary conditions;
\begin{align}
 - \left. \partial_r C_o \right|_{r=1} &   = \quad \frac{D_{hp}}{2D_o}   \mathcal{K}_{\text{eff}}^{(hp)} \Big(1 + C_{hp} \Big) K(\cos \theta)  , \ \\
- \left. \partial_r C_{hp} \right|_{r=1} & = - \  \mathcal{K}_{\text{eff}}^{(hp)} \Big( 1 + C_{hp} \Big) K(\cos \theta)  \ . 
\end{align}
This gives rise to a system of equations :
\begin{equation}
\sum_{l=0}^{\infty} W_l (l+1) P_l(\cos\theta) = \mathcal{K}_{\text{eff}}^{(hp)} \left( 1 - \sum_{l=0}^{\infty} W_l P_l(\cos\theta) \right) K(\cos\theta) \ .
\end{equation}
From this, using the orthogonality condition of the Legendre polynomials, we obtain the linear system of equations for the amplitudes, $W_l$:
\beq
\mathbb{M}^{(d)} \cdot \mathbf{W} = {\bs \Lambda}^{(d)} \; ,
\eeq
where $\mathbf{W} = ( W_0, \ldots , W_l , \ldots )$, and more explicitly
\begin{equation}
\begin{pmatrix}
1  &    0  &    0   &  \cdots &     0  & \cdots\\
0  & M_{11}^{(d)}& M_{12}^{(d)}&  \cdots & M_{1l}^{(d)}& \cdots\\
0  & M_{21}^{(d)}& M_{22}^{(d)}&  \cdots & M_{2l}^{(d)}& \cdots\\
\vdots  & \vdots& \vdots& \ddots & \vdots& \hdots\\
0  & M_{n1}^{(d)}& M_{n2}^{(d)}&  \hdots & M_{nl}^{(d)}& \hdots\\
\vdots& \vdots& \vdots& \vdots& \vdots&  \ddots
\end{pmatrix}
\begin{pmatrix}
 W_0 \\ W_1\\ W_2\\ \vdots\\ W_l \\  \vdots
\end{pmatrix}
= \begin{pmatrix}
0 \\ \Lambda_1^{(d)}\\ \Lambda_2^{(d)}\\  \vdots \\ \Lambda_n^{(d)} \\ \vdots
\end{pmatrix}  \label{solution:Wls}
\end{equation}
where the matrix $\mathbb{M}^{(d)}$ and vector $\bs{\Lambda}^{(d)}$ entries are given by 
\begin{align}
M_{nl}^{(d)} & = \delta_{nl} +  \mathcal{K}_{\text{eff}}^{(hp)} \left( \frac{n + \frac{1}{2}}{n + 1 }\right) \int_0^1 P_n(x)  P_l(x)  \ dx \  , \\ 
\Lambda_n^{(d)} & = \mathcal{K}_{\text{eff}}^{(hp)}   \left( \frac{ n + \frac{1}{2} }{n + 1 }\right) \int_0^1 P_n(x) \ dx \  .
\end{align}
Here as in the ionic  section, we solve a truncated approximation of the linear equations above, including all modes up to the $N$'th Legendre mode$\{ W_0, W_1, \cdots , W_N \}$. As above, we can obtain analytic asymptotic solutions for $\mathcal{K}_{\text{eff}}^{(hp)} \ll 1$ and $\mathcal{K}_{\text{eff}}^{(hp)} \gg 1$: }\\

\par \noindent \tl{$\mathcal{K}_{\text{eff}}^{(hp)} \ll 1:$ In this regime, $W_l \sim \Lambda_l^{(d)}$ and 
\begin{equation}
W_1 \sim \frac{3}{8} \mathcal{K}_{\text{eff}}^{(hp)}
\end{equation} }
\par \noindent $\mathcal{K}_{\text{eff}}^{(hp)}  \gg 1:$ In this regime, 
\begin{equation}
W_1 \sim   \Xi
\end{equation}
where $\Xi > 0$ is some constant to be determined numerically. Since $\mathcal{K}_{\text{eff}}^{(hp)} \propto a$ (swimmer size), this implies the limit $\mathcal{K}_{\text{eff}}^{(hp)} \gg 1$ corresponds to large swimmer size.
\tbl{For $a=1.00\mu$m sized swimmer in $10$\% w/v H$_2$O$_2$ solution, and the measured reaction rates in table (\ref{system:parameters}), the estimate of the dimensionless reaction rate coefficient is $\mathcal{K}_{\text{eff}}^{(hp)} \approx 0.026$. Hence, this puts the current experimental measurements~\citep{Ebbens2014,Brown2014} in the first regime $(W_l \sim \Lambda_l^{(d)} )$. We note that in this regime $C_{hp} \sim W_l \sim \mathcal{K}_{\text{eff}}^{(hp)} \ll 1$.}
%\begin{equation}
%W_l =  \left( \frac{l+\frac{1}{2}}{l + 1} \right) \left( \frac{ \mathcal{K}_{\text{eff}}^{(hp)}  a}{D_{hp} c_{hp}^{\infty}} \right) \int_0^1 P_l(x) dx
%\end{equation}
%%

\tl{The coefficients $W_l$, determine the solute concentration, and hence the slip velocity which act as boundary conditions for the Stokes flow problem. Hence the velocity fields generated, expressed in terms of the fundamental singularities (see equation (\ref{eq:stokes_sing})) of Stokes flow are 
\begin{align}
\bs{v}(\bs{r}) & =    B_1^{(d)} \ \bs{D}(\bs{r})   +  B_3^{(d)} \ \left[ \partial_z^2 \bs{G} (\bs{r})\right]  + \ \mathcal{O}(r^{-4}) \ , \label{flow:free:space:solution}
\end{align} 
where the coefficients ($B_1^{(d)},B_3^{(d)}$) are
\begin{equation}
B_1^{(d)} = - \frac{1}{3} \mu_d W_1 \left( 1 - \frac{3}{2} \frac{W_3}{W_1}  \right) ; \quad B_3^{(d)} =  \frac{5}{4} \mu_d W_3 \ . 
\end{equation} }

\iffalse
\par Therefore, the gradients generated by asymmetric catalytic activity on the swimmer surface contributes the diffusiophoretic propulsion velocity 
\begin{equation}
\mathbf{U}^d = - \sum_{i \in \{hp,o \}} \frac{1}{4\pi} \int_0^{2\pi} d \phi \int_0^{\pi} \sin \theta \ d \theta \ \mu_d^{(i)}  \left( \mathds{1} - \bs{\hat{n} \hat{ n}} \right) \cdot \nabla C_i  \ ,
\end{equation}
\fi
%
%Recall from our non-dimensionalisation, that lengths are measured in units of $a$ and $\mathbf{U}^d $ is measured in units of $c_{hp}^{\infty} k_B T L^2/a \ \eta $. 
%
Imposing the condition of  net zero total force,  we obtain the the neutral diffusiophoretic contribution to the propulsion velocity as
\begin{equation}
\mathbf{U}^d = - \sum_{i \in \{hp,o \}} \frac{1}{4\pi} \int_0^{2\pi} d \phi \int_0^{\pi} \sin \theta \ d \theta \ \mu_d^{(i)}  \left( \mathds{1} - \bs{\hat{n} \hat{ n}} \right) \cdot \nabla C_i  \ ,
\end{equation}
where since we have taken the interaction potential, $\Psi$ identical for all species, we have identical neutral diffusiophoretic mobilities for all the neutral solute species, $ \mu_d^{(i)} =1 \; , \; i \in \{ o,hp\}$. From the modes calculated above, we thus obtain
\begin{equation}
\mathbf{U}^d = - \frac{2}{3} \ \mu_d \ W_1   \ \bs{\hat{e}}_z \ , \label{diffusio:contr}
\end{equation}
where $\mu_d = \mu_d^{(hp)} - (D_{hp}/2D_o) \mu_d^{(o)} =1 - (D_{hp}/2D_o) $ is the combined effective diffusiophoretic mobility.
%
%%%%%%%%%%%%%%%%%%%%%%%%%%
% Below are the additional citations to the pH dependence of the H2O2 decomposition
%%%%%%%%%%%%%%%%%%%%%%%%%%%
\iffalse
@article{liu2014ph,
  title={pH dependent catalytic activities of platinum nanoparticles with respect to the decomposition of hydrogen peroxide and scavenging of superoxide and singlet oxygen},
  author={Liu, Yi and Wu, Haohao and Li, Meng and Yin, Jun-Jie and Nie, Zhihong},
  journal={Nanoscale},
  volume={6},
  number={20},
  pages={11904--11910},
  year={2014},
  publisher={Royal Society of Chemistry}
}
%
%
@article{mckee1969catalytic,
  title={Catalytic decomposition of hydrogen peroxide by metals and alloys of the platinum group},
  author={McKee, DW},
  journal={Journal of Catalysis},
  volume={14},
  number={4},
  pages={355--364},
  year={1969},
  publisher={Elsevier}
}
\fi
%
%%%%%%%%%%%%%%%%%%%%%%%%%%%%%
\subsection{Comparison of ionic and neutral velocities}
\begin{figure}
\begin{center}
\includegraphics[scale=.4]{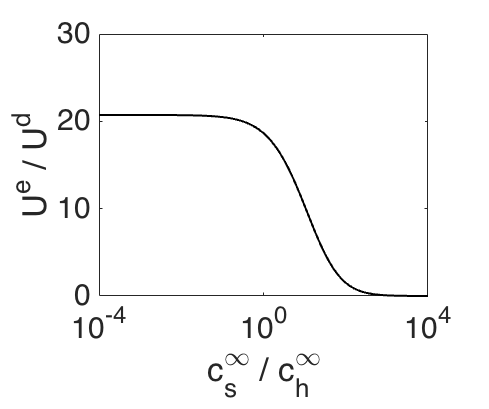} 
\caption{\yhy{Plot of comparison of the ionic solutes contribution to the neutral solutes contribution ( eqn. \ref{U_ratio}) and system parameters in table \ref{system:parameters} and $\text{pH}=5.8$.  }}
\label{fig:U_comparison}
\end{center} 
\end{figure}

\tl{Finally, we can now compare the two contributions to the swimmer propulsion from ionic and neutral solutes using dimensional quantities. From equations (\ref{electro:contr},\ref{diffusio:contr}), the relative speed 
\begin{equation}
\frac{\bar{U}^e}{\bar{U}^d}  \ = \  \frac{ \left(\epsilon k_B^2T^2/a \eta e^2 \right) U^e }{  \left(\bar{\mu}_d^{\ddagger} c_{hp}^{\infty}/a \right) U^d }  \  =  \ \frac{\bar{\mu}_e }{\bar{\mu}_d} \frac{(k_BT/e) \gamma^{(1)} A_1 }{ c_{hp}^{\infty} W_1}  \ ,
\end{equation} 
where $\bar{\mu}_e = (\epsilon k_BT/e\eta ) \, \mu_e$ is the electrophoretic mobility and $\bar{\mu}_d = \mu_d \bar{\mu}_d^{\ddagger} \, $ is the diffusiophoretic mobility both in dimensional form. 
% is the dimensional mobilities.
For a fixed swimmer size, and in the limit $\gamma^{(0)} \ll 1, \ \mathcal{K}_{\text{eff}}^{(hp)} \ll 1$, the above ratio takes the simple analytic expression 
\begin{equation}
\frac{\bar{U}^e}{\bar{U}^d} = \frac{1}{6} \frac{\bar{\mu}_e}{\bar{\mu}_d} \frac{D_{hp}}{D_h} \frac{\Delta k_{\text{eff}}^{(h)}}{k_{\text{eff}}^{(hp)} } \frac{\left( k_BT/e\right)}{(c_s^{\infty} + c_h^{\infty})} \ ,  \label{U_ratio}
\end{equation} 
where $k_{\text{eff}}^{(hp)} c_{hp}^{\infty}>0$ is the effective rate of the hydrogen peroxide consumption and  $\Delta k_{\text{eff}}^{(h)} >0$ is the scale of the {difference} between the rates at the pole and equator  due to the Pt thickness variation (defined in Appendix \ref{appA} for a particular example of reaction model). These rates are linear functions of the $c_{hp}^{\infty}$ concentration for low fuel concentration. In Fig. (\ref{fig:U_comparison}), it can be seen that the electrophoretic contribution vanishes at large ionic strengths, and the swimmer speed asymptotically approaches the diffusiophoretic contribution value $\bar{U}^d$. The self-diffusiophoretic speed $\bar{U}^d = \bar{\mu}_d k_{\text{eff}}^{(hp)} c_{hp}^{\infty}/4 D_{hp} \sim 0.52 \mu \mbox{ms}^{-1}$ (see table \ref{system:parameters}) for the chosen system parameter values in the plot (Fig. \ref{fig:U_comparison}).}

\section{Summary and discussion}

\tl{Therefore, the total propulsion velocity of the metallic-insulator sphere from both electrophoresis and diffusiophoresis, from equations (\ref{electro:contr} and \ref{diffusio:contr}), in dimensional form is 
\begin{equation} 
\bar{\mathbf{U}} =  - \left[  \frac{1}{3} \  \bar{\mu}_e   \frac{ \ \Delta k_{\text{eff}}^{(h)} c_{hp}^{\infty}}{D_h \left( c_{s}^{\infty} + c_h^{\infty}\right)} \left( \frac{k_BT}{e} \right) A_1  + \frac{2}{3} \  \bar{\mu}_d \frac{ c_{hp}^{\infty}}{a}  W_1  \right] \hat{\bs{e}}_z  \ ,
\end{equation}  
\tl{where $\bar{\mu}_e $ and $\bar{\mu}_d$ are the electrophoretic and diffusiophoretic mobilities. 
%The effective rate of the hydrogen peroxide consumption ${\cal K}_{\text{eff}}^{(hp)}$ and 
The scale of the difference between the rates at the poles and equator due to the Pt thickness variation $\Delta k_{\text{eff}}^{(h)}>0$ is defined in {Appendix} \ref{appA} for a particular  reaction kinetic model. }
%$\Delta k_{\text{eff}}^{(h)} >0$ is the scale of the {difference} between the rates at the pole and equator  due to the Pt thickness variation
%These rates are linear functions of the $c_{hp}^{\infty}$ concentration for low fuel concentration.
We point out that these results are qualitatively independent of the details of the reaction kinetics, provided the reaction involves both charged and neutral pathways for the reduction of the hydrogen peroxide and the reaction rate varies along the catalytic cap. }

The \yhy{ionic} contribution to the expression above has a number of important simple features that are in agreement with recent experimental results on this system \cite{Howse2007,Ebbens2012,Brown2014,Ebbens2014,Das2015}: 
{(1)}
 it depends linearly on the fuel, $c_{hp}^\infty$ at low concentrations and the dependence weakens at high concentrations, 
 {(2)} 
 it is independent of $a$ at small $a$ and behaves as $1/a$ for large $a$ due to fuel depletion as shown in Ref. \cite{Ebbens2012}, and 
 {(3)} 
 it is a monotonically decreasing function of salt concentration, $c_s^\infty$ starting from a finite value when $c_s^\infty=0$ and tending to zero as $c_s^\infty$ becomes large. Hence at high salt concentration the swimming speed saturates to the neutral diffusiophoretic value \yhy{(see Fig. \ref{fig:U_comparison})}.
The electrophoretic contribution, which can be much larger than the diffusiophoretic part, vanishes if there is no variation in the rates $k_i$ on the surface.

%\subsection{Salt effects}
Adding salt to the solution containing the swimmer would influence the propulsion in three possible ways (1) pH neutral salts that do not specifically adsorb to the surface would enhance the solution conductivity thereby reducing the effective screening length (2) while alkali or acidic salts would in general alter the total surface charge in addition to the increased solution conductivity (3) Pt catalytic decomposition of $\text{H}_2\text{O}_2$ is known to strongly depend on the solution pH~\citep{liu2014ph,mckee1969catalytic}. Hence, non-pH neutral salts would also affect the Pt catalytic activity.
%This is likely due to the complex reaction kinetics with $[\text{H}^+]$ ions acting as non-competetive inhibitors to the active catalytic sites.

We note also that due to the existence of the two separate reaction loops, the overall catalytic reaction rate (measured from the current $J_o$ above) can be significantly reduced with only small reductions to the swimming speed; say by a significant decrease in $k_1$. This type of behaviour would be expected from any reaction scheme which has this topological structure.

In conclusion, we have shown that in a system with catalytic reaction with charged intermediates, the existence of thickness-dependence in the reaction rates up to a certain limit (a few nanometres), allows us to create---by tapering the catalyst layer---spatially separated nonequilibrium cycles that could lead to large scale (many microns) ionic currents in the form of closed loops in the bulk. This remarkable effect, combining long range electrostatic interactions with nonequilibrium chemical reactions to substantially enhance surface generated flows has potential for application in many different areas of nanoscience.

%\section{Conclusion}

\section{Acknowledgments}
 This work was supported by EPSRC grant EP/G026440/1 (TBL, YI), and HFSP grant RGP0061/2013 (RG). YI acknowledges the support of University of Bristol. TBL acknowledges support of BrisSynBio, a BBSRC/EPSRC Advanced Synthetic Biology
 Research Centre (grant number BB/L01386X/1).

\appendix

\section{Reaction kinetics}\label{appA}
\tbl{In this section, our goal is to obtain the fluxes on the surface of the swimmer,  $\bar{\mathcal{J}}_i$ of all the chemical species involved in the H$_2$O$_2$ decomposition
\begin{equation}
\mbox{Pt} +  2{\text{H}_2\text{O}_2} \ \rightarrow \boxed{\textsf{Intermediate complexes}} \rightarrow \ \mbox{Pt} + 2\text{H}_2\text{O} + \text{O}_2 \ . \label{h2o2:decomp:generic:app}
\end{equation} 
Though a complete picture of the intermediate complexes in reaction (\ref{h2o2:decomp:generic:app}) remains elusive, it is known that 
there are neutral pathways as well as ionic electrochemical pathways~\citep{Hall2000,Katsounaros2012}. However, we find that our results are qualitatively  independent of many details of the reaction scheme considered as long as they involve both neutral and charged pathways. So our lack of knowledge of 
the microscopic chemical kinetics is not such a hindrance.
To illustrate this, we consider two different reaction schemes  %(\ref{h2o2:decomp:generic:app}) 
involving a neutral as well a charged pathway. We emphasize that both schemes are provided simply as examples as the precise details of the chemical kinetics are not known.}

\subsection{Reaction scheme 1}
\tbl{First, we consider a reaction scheme for the reaction (\ref{h2o2:decomp:generic:app}) made up of two pathways, one neutral 
%made up of two pathways, one neutral 
% taking into account both the inert chemical pathway
\begin{equation}
\mbox{Pt} +  2{\text{H}_2\text{O}_2} \ \overset{k_0}{\longrightarrow} \ \mbox{Pt}\left({\text{H}_2\text{O}_2} \right)  + {\text{H}_2\text{O}_2} \ \overset{k_1}{\rightarrow }\ \mbox{Pt}\left({\text{H}_2\text{O}_2} \right)_2 \ \overset{k_2}{\rightarrow} \  \mbox{Pt} + 2\text{H}_2\text{O} + \text{O}_2 \ ,  \label{h2o2:decomp:neutral}
\end{equation} 
and the other ionic involving charged intermediates,
%and an electrochemical pathway involving charged intermediates
\begin{align}
\mbox{Pt} +  2{\text{H}_2\text{O}} \quad  \underset{k_{-3}}{\overset{k_3}{\rightleftharpoons}} \quad \mbox{Pt}\left({\text{H}_2\text{O}_2} \right) + 2e^-  + 2\text{H}^+ \ . \label{h2o2:decomp:charged}
\end{align} }
%In addition, water naturally dissociate into protons and hydroxide ions
%\begin{equation}
%{\text{H}_2\text{O}} \ \rightleftharpoons \ 2\text{H}^+ + \text{OH}^-.
%\end{equation}

\tl{The reaction scheme above and the intermediate states denoted by $(0,1,2)$ are enumerated in Fig. (\ref{reaction_scheme}).
\begin{figure}
\begin{center}
\includegraphics[scale=.4]{reaction_schematic1}
\end{center}
\caption{(\tbl{\textbf{Reaction scheme 1}): Schematic complexation kinetics of the Platinum catalyst with free ($0$'th state) Pt occupied with probability density $p_0$; first complex state $\text{Pt}(\text{H}_2\text{O}_2)$ occupied with probability density $p_1$, and the second complex state $\text{Pt}(\text{H}_2\text{O}_2)_2$ occupied with probability density $p_2$.}} 
\label{reaction_scheme}
\end{figure}
The kinetics of the Pt catalyst complexation in stationary state reads
\begin{align}  
0 & = \partial_t p_0 = -\  k_0  \bar{c}_{hp} \ p_0 - k_3 \ p_0 + k_2 \ p_2  + k_{-3} \bar{c}_h^2 \ p_1 \ , \\
0 & = \partial_t p_1 =   \quad k_0  \bar{c}_{hp} \ p_0 + k_3 \ p_0 - k_1 \bar{c}_{hp} \ p_1  - k_{-3}  \bar{c}_h^2 \ p_1  \  , \\
0 & = \partial_t p_2 = \quad k_1  \bar{c}_{hp} \ p_1 - k_2  \ p_2 \  , \label{probability:kinetics:inhomogeneous} 
\end{align}
where $p_i$'s are the complexation probabilities. Solving for these probabilities $p_i$, we obtain
\begin{align}
p_0 & = \mathcal{M}^{-1} k_2 \left( k_1 \bar{c}_{hp} + k_{-3} \bar{c}_h^2  \right)  \  ,  \\
p_1 & = \mathcal{M}^{-1} k_2 \left( k_0 \bar{c}_{hp} + k_3 \right)  \ ,   \\
p_2 & = \mathcal{M}^{-1} k_1 \bar{c}_{hp} \left( k_0  \bar{c}_{hp} + k_3  \right)  \ , 
\end{align}
where the normalization condition $p_0+p_1+p_2=1$ was used and we have defined
\begin{equation}
\mathcal{M} := k_2 k_3 + k_2 k_{-3}  \bar{c}_h^2 +  \left( k_0 k_2 + k_1 k_3 + k_1 k_2 \right)  \bar{c}_{hp} + k_0 k_1  \bar{c}_{hp}^2   \ .
\end{equation}
This  leads to expressions for the fluxes $\bar{\mathcal{J}}_i$ of $i \in \{o,hp,h\}$
\begin{align}
 \bar{\mathcal{J}}_o(\theta)   &= \quad k_2 \ p_2 K(\cos \theta) \ , \nonumber \\
 &  =  \mathcal{M}^{-1} k_1 k_2 \bar{c}_{hp} \Big( k_0  \bar{c}_{hp} + k_3  \Big) K(\cos\theta) \ , \label{o2:flux} \\
 \bar{\mathcal{J}}_{hp}(\theta)&=  -  \left( k_0 \ p_0 + k_1 \ p_1  \right) \bar{c}_{hp}  K(\cos \theta) \ , \nonumber  \\
 & = -  \mathcal{M}^{-1} k_2\bar{c}_{hp} \Big( k_1k_3 + k_0k_{-3}\bar{c}_h^2 + 2 k_0k_1\bar{c}_{hp}  \Big) K(\cos\theta) \ , \label{hp:flux}  \\
\bar{\mathcal{J}}_{h}(\theta)  &=  2 \ \left( k_3 \ p_0 - k_{-3} \bar{c}_h^2 \ p_1  \right)  K(\cos \theta)  \ , \nonumber  \\
& = 4 \mathcal{M}^{-1} k_2 \bar{c}_{hp} \Big( k_1k_3 - k_0k_{-3} \bar{c}_h^2 \Big) K(\cos\theta) \ ,  \label{h:flux}
\end{align} 
and for the fluxes of hydroxide and the salt, $\bar{\mathcal{J}}_{oh} = 0, \bar{\mathcal{J}}_{s,\pm} = 0$. Measurements of the reaction rates~\citep{Ebbens2014} imply that $\bar{\mathcal{J}}_{hp}$ and $\bar{\mathcal{J}}_o$ vary with the Pt coating thickness (of $\sim$nm scale). Hence we may assume that the rate `constants' $k_i$'s vary in a similar manner.  }  \\

\tl{Since the thickness of the coating varies across the Pt cap,  the reaction rates  $k_i(\theta)$'s vary over the coated hemisphere, and can be expanded in Legendre polynomials. We consider a simple linear approximation
\begin{equation}
k_i(\theta) \cong k_i^{(0)} + k_i^{(1)} \cos \theta \  ,
\end{equation}
in $\cos\theta$ and we assume weak variation $k_i^{(1)}/k_i^{(0)} \ll 1$ of the rates.}  \\

%(\emph{Proton flux})  
\tl{The solute fluxes $\bar{\mathcal{J}}_i$ above in eqns. (\ref{o2:flux} - \ref{h:flux}) require the inner (Debye-layer) proton concentration profile
\begin{equation}
\bar{c}_h(1,\theta)/c_h^{\infty} =  \left( 1 + C_h(1,\theta) \right) e^{- \zeta(\theta)} \ ;  \qquad  \zeta(\theta) = \varphi_s -  \Phi(1,\theta) \ ,
\end{equation}
from eqn. (\ref{inner:outer:relation}) in Appendix \ref{appB}, where $C_h(1,\theta)$ and $\Phi(1,\theta)$ are the deviations from the uniform background of proton concentration and electric fields. $\varphi_s$ is the electric potential on the   swimmer surface. Hence, the proton flux at the outer edge of the double-layer reads
\begin{equation}
\bar{\mathcal{J}}_h ( \theta ) =  4 \mathcal{M}^{-1} k_2 \bar{c}_{hp} \  \left( k_1 k_3 - k_0 k_{-3} \left(c_h^{\infty}\right)^2 (1+C_h)^2 e^{- 2 \zeta(\theta) }\right)  K(\cos \theta) \  .
\end{equation} }

\tl{Furthermore, Taylor-expanding the flux up to linear order in $k_i^{(1)}$, and the deviations $C_h, \ \Phi$ ;
\begin{equation} 
\bar{\mathcal{J}_h} (\theta) = \left( \bar{\mathcal{J}}_h^{(0)}  \ + \ \bar{\mathcal{J}}_h^{(1)} \ P_1(\cos \theta)  + \ \bar{\mathcal{J}}_h^{'(0)} \  \left(\Phi + C_h \right)  \right) K(\cos\theta)   %+  \mathcal{O}\left(\frac{ \bar{\mathcal{J}}_h^{(j)} \mathcal{M}^{(1)}}{\mathcal{M}^{(0)}}\right) 
\ , \label{H:flux:approx0}
\end{equation} 
where $j= \{ 0,1,2\}$. We define 
\begin{align}
\bar{\mathcal{J}}_h^{(0)} & =  4  \{ k_2 c_{hp} \}^{(0)} \ \left( \{k_1k_3\}^{(0)} - \{k_0k_{-3}\}^{(0)} (c_h^{\infty})^2 e^{-2 \varphi_s } \right)/\mathcal{M}^{(0)} \ ,  \label{H:flux:0} 
 \\
\bar{\mathcal{J}}_h^{(1)} & =  4 \{ k_2 c_{hp} \}^{(0)} \ \left( \{k_1k_3\}^{(1)} - \{k_0k_{-3}\}^{(1)} (c_h^{\infty})^2 e^{-2 \varphi_s  } \right) / \mathcal{M}^{(0)} \ ,
\label{H:flux:1}  
\\
\bar{\mathcal{J}}_h^{'(0)} & =  8 \{ k_2 c_{hp} \}^{(0)} \{k_0k_{-3}\}^{(0)} (c_h^{\infty})^2 e^{-2 \varphi_s }  / \mathcal{M}^{(0)} \ , \label{H:flux:2}  \\
\mathcal{M}^{(0)} & = \left \{ k_2 k_3 + k_2 k_{-3}  \bar{c}_h^2 +  \left( k_0 k_2 + k_1 k_3 + k_1 k_2 \right)  \bar{c}_{hp} + k_0 k_1  \bar{c}_{hp}^2 \right\}^{(0)}  \ . 
\end{align} 
Now, imposing net charge conservation, equation (\ref{quasi:steady:constraint}) on the swimmer surface,  equation (\ref{H:flux:approx0}) for the proton flux leads to 
\begin{equation}
\bar{\mathcal{J}}_h^{(0)} + \frac{1}{2} \bar{\mathcal{J}}_h^{(1)} + \bar{\mathcal{J}}_h^{'(0)} \int_0^{\pi}  \left( \Phi + C_h\right)  K(\cos \theta) \ \sin \theta \ d \theta  = 0 \ .
\end{equation}
Then, substituting for $\bar{\mathcal{J}}_h^{(i)}$'s (from eqns. \ref{H:flux:0}-\ref{H:flux:2}) and simplifying ;
\begin{align}  
    (c_h^{\infty})^2 e^{-2 \varphi_s }  \   &  \left(   1 + \frac{1}{2} \frac{ \{k_0k_{-3}\}^{(1)}}{\{k_0k_{-3}\}^{(0)} }    +  2  \int_0^{\pi} \left( \Phi + C_h \right)K(\cos\theta) \sin\theta \, d\theta  \right)   \\
   &\qquad \qquad\qquad \qquad \qquad  \qquad  = \quad   \frac{\{k_1k_{3}\}^{(0)}}{\{k_0k_{-3}\}^{(0)} }    + \frac{1}{2}  \frac{\{k_1k_{3}\}^{(1)}}{\{k_0k_{-3}\}^{(0)} }    \     . 
\end{align} 
For a uniform coating (i.e $k_i^{(1)} = 0$ for all $i$'s), which has the trivial solution $\Phi = C_h =  0$, the zero total current condition gives rise to 
% Remember, we are expanding about the reference uniform catalyst coating surface state (i.e $k_i^{(1)} = 0$ for all $i$'s); which has the trivial 
% solution $\Phi = C_h =  0$ . This implies that to the leading linear order of our expansion, the steady state constraint of vanishing 
%(surface average) proton current is satisfied by
\begin{align}  
  (c_h^{\infty})^2 e^{-2 \varphi_s }  \   \cong  \    \frac{\{k_1k_{3}\}^{(0)}}{\{k_0k_{-3}\}^{(0)} }    \   .   \label{H:balance}
\end{align} 
which is all that is required for a linear expansion about a uniform coating.
Hence solving for the swimmer potential $\varphi_s$, we obtain the equation 
\begin{equation}
\varphi_s  = \varphi_s^{\ominus} - \ln 10 \ \mbox{pH} \ , \label{zeta:potential:equilibrium}
\end{equation}
where $\mbox{pH} = - \log_{10} c_h^{\infty}$ (with $c_h^{\infty}$ measured in molar units) and 
%the kinetically defined potential
\begin{equation}
 \varphi_s^{\ominus} =  - \frac{1}{2}  \ln \left(  \frac{ \{ k_1 k_3 \}^{(0)}  }{  \{ k_0 k_{-3} \}^{(0)} } \right) \ .
\end{equation} 
Therefore, eliminating the swimmer potential and proton background concentration ($\varphi_s, \, c_h^{\infty}$) by substituting eqn. (\ref{H:balance}) into eqn. (\ref{H:flux:approx0}) and keeping only linear perturbations, the proton flux assumes a simple form
\begin{equation} \boxed{
\bar{\mathcal{J}}_h(\theta)  \cong \left[ \Delta k_{\text{eff}}^{(h)} \  \left( 1 - 2 \cos \theta \right)   - k_{\text{eff}}^{(h)} \ \delta \left(\Phi + C_h \right) \right] c_{hp}^{\infty} \  K(\cos \theta)  \ , } \label{H:flux:approximate}
\end{equation}
where we have defined 
\begin{align}
k_{\text{eff}}^{(h)} & = \frac{8 \{k_1 k_2k_3\}^{(0)}}{\mathcal{M}^{(0)} } \ ; \label{gamma:0:definition} \\  
\Delta k_{\text{eff}}^{(h)}  & = \frac{ k_{\text{eff}}^{(h)} }{2} \left( \frac{\{ k_0 k_{-3} \}^{(1)}  }{  \{k_0 k_{-3} \}^{(0)} }  - \frac{\{k_1 k_3 \}^{(1)}}{\{ k_1 k_3 \}^{(0)}} \right) \ ,  \label{gamma:1:definition} \\
\delta \left(\Phi + C_h \right) & = \left[ (\Phi + C_h) - \int_0^{\pi}  (\Phi + C_h) K(\cos\theta) \ \sin \theta \ d  \theta \right] \ . \label{eq:deviations}
\end{align} 
\tl{$k_{\text{eff}}^{(h)} c_{hp}^{\infty}>0$ is the typical scale  of the average proton consumption and production} and $\Delta k_{\text{eff}}^{(h)} c_{hp}^{\infty}$ \tl{the scale of the \emph{difference} between the rates at the pole and equator} due to variation of coating thickness over the surface.  
$\delta\left(\Phi + C_h\right)$ is the deviation of the perturbative fields from their surface average; which promotes/penalise the oxidation/reduction reactions. 
The proton flux $\bar{\mathcal{J}}_h$ is linear in $c_{hp}^{\infty}$ for low fuel concentration and the dependence weakens for high fuel concentration.
It is noteworthy that with uniform coating, $k_i = k_i^{(0)} \Rightarrow \Delta k_{\text{eff}}^{(h)} =0$ and the deviation fields vanish ($\Phi = 0 = C_i$). } \\
%$k_{\text{eff}}^{(h)} >0 $ is the typical scale  of the average proton consumption and production while $\Delta k_{\text{eff}}^{(h)} >0$ is the scale of the \emph{difference} between the rates at the pole and equator

%(\emph{Neutral solutes}) 
\tl{The fluxes of neutral solutes from eqns. (\ref{o2:flux},\ref{hp:flux}) give 
\begin{align}
\bar{\mathcal{J}}_o(\theta)  \cong \  \frac{1}{2} \ k_{\text{eff}}^{(hp)} \bar{c}_{hp}(1,\theta) \ , \\
\bar{\mathcal{J}}_{hp}(\theta) \cong \  - \  k_{\text{eff}}^{(hp)}  \bar{c}_{hp}(1,\theta)  \ ,
\end{align}
%\begin{align}
% - \left. \frac{\partial C_o }{ \partial r} \right|_{r=1} & = \mathcal{J}_{o}(\theta) K(\cos \theta) \cong \quad \frac{1}{2} \left( \frac{ \mathcal{K}_{\text{eff}}^{(hp)}  a}{D_{o} c_{hp}^{\infty}} \right) K(\cos \theta) \ , \\
%- \left. \frac{\partial C_{hp} }{ \partial r} \right|_{r=1} & =  \mathcal{J}_{hp}(\theta) K(\cos \theta) \cong - \left( \frac{ \mathcal{K}_{\text{eff}}^{(hp)}  a}{D_{hp} c_{hp}^{\infty}} \right) K(\cos \theta) \ ,
%\end{align}
where the effective rate of hydrogen peroxide consumption is defined 
\begin{equation}
k_{\text{eff}}^{(hp)} = \frac{ 2}{\mathcal{M}^{(0)}} \{k_0 k_1 k_2 \}^{(0)} \bar{c}_{hp}(1,\theta) + \frac{1}{2} k_{\text{eff}}^{(h)} \ ,
\end{equation}
and the effective rate of proton consumption/desorption $k_{\text{eff}}^{(h)}$ is defined in equation (\ref{gamma:0:definition}). Note that $\bar{\mathcal{J}}_{hp}$ and $\bar{\mathcal{J}}_{o}$ are linear in $c_{hp}^{\infty}$ for low fuel ($c_{hp}^{\infty}$) concentration, and show the saturation typical of Michaelis-Menten kinetics at high fuel concentration.  }
% Note that $D_{hp} \mathcal{J}_{hp} + 2 D_o \mathcal{J}_o = 0$ and hence it suffices to solve for either of the fields to obtain the other.

\subsection{Reaction scheme 2}
\tbl{Alternatively, we may consider a different reaction scheme, with the same neutral pathway 
\begin{equation}
\mbox{Pt} +  2{\text{H}_2\text{O}_2} \ \overset{k_0}{\longrightarrow} \ \mbox{Pt}\left({\text{H}_2\text{O}_2} \right)  + {\text{H}_2\text{O}_2} \ \overset{k_1}{\rightarrow }\ \mbox{Pt}\left({\text{H}_2\text{O}_2} \right)_2 \ \overset{k_2}{\rightarrow} \  \mbox{Pt} + 2\text{H}_2\text{O} + \text{O}_2 \ ,  \label{h2o2:decomp:neutral2}
\end{equation} 
but with a different electrochemical pathway 
\begin{align}
& \mbox{Pt}\left({\text{H}_2\text{O}_2} \right) \quad \overset{k_3}{\longrightarrow} \quad \mbox{Pt} +  2\text{H}^+  + 2e^- + \text{O}_2 \ , \label{h2o2:decomp:oxidation}  \\ 
& \mbox{Pt}\left({\text{H}_2\text{O}_2} \right) + 2\text{H}^+  + 2e^- \quad \overset{k_4}{\longrightarrow} \quad \mbox{Pt} +  2\text{H}_2\text{O} \ .
\end{align} 
This is the reaction scheme commonly used in modeling the electrophoretic motion of the bimetallic nanorods~\citep{Paxton2005,Dhar2006,Kline2005,Kline2005b,Sabass2012b}.
As for the reaction scheme considered in the previous section, we can write down the equations of motion for the kinetics for this  scheme (see Fig.~\ref{reaction_scheme2}).
\begin{figure}
\begin{center}
\includegraphics[scale=.4]{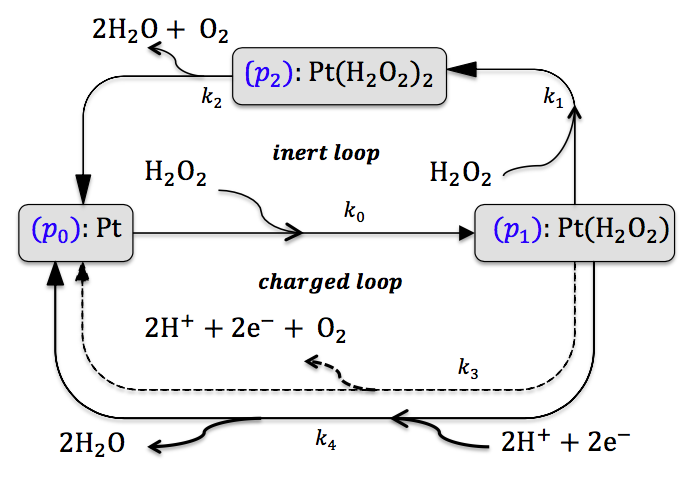}
\end{center}
\caption{(\textbf{Reaction scheme 2}): Schematic complexation kinetics of the Platinum catalyst with free ($0$'th state) Pt occupied with probability density $p_0$; first complex state $\text{Pt}(\text{H}_2\text{O}_2)$ occupied with probability density $p_1$, and the second complex state $\text{Pt}(\text{H}_2\text{O}_2)_2$ occupied with probability density $p_2$.} 
\label{reaction_scheme2}
\end{figure}
Hence, we can, as in the previous section,  obtain the fluxes $\bar{\mathcal{J}}_i$'s, 
%\begin{align}
%p_0 & =  \mathcal{M}^{-1} \, k_2 \left( k_1 \bar{c}_{hp} + k_3 + k_4 \bar{c}_{h}^2 \right) \  , \\
%p_1 & = \mathcal{M}^{-1} \, k_0 k_2 \bar{c}_{hp}  \ , \\
%p_2 & = \mathcal{M}^{-1}  \, k_0 k_1 \bar{c}_{hp}^2  \  , 
%\end{align} 
\begin{align}
\bar{\mathcal{J}}_{h}(\theta) & =  2 \, \mathcal{M}^{-1} \, k_0 k_2 \bar{c}_{hp} \left( k_3 - k_4 \bar{c}_{h}^2 \right) \, K(\cos \theta)   \  ,  \\
\bar{\mathcal{J}}_{hp}(\theta) & = - \mathcal{M}^{-1} \, k_0 k_2 \bar{c}_{hp} \left( 2k_1 \bar{c}_{hp} + k_3 + k_4 \bar{c}_{h}^2  \right) \,  K(\cos \theta)  \   , \\
\bar{\mathcal{J}}_o (\theta) & = \mathcal{M}^{-1} \, k_0 k_2 \bar{c}_{hp} \left( k_1 \bar{c}_{hp} + k_3 \right) \, K(\cos \theta)  \  ,
\end{align} 
where here
$\mathcal{M}  : = k_2 k_3 + \left( k_0 + k_1 \right)k_2 \bar{c}_{hp}  + k_2k_4\bar{c}_{h}^2 + k_0 k_1 \bar{c}_{hp}^2 \  .$
} \\

\par \noindent %\textit{(Protons flux)} 
\tbl{Now, imposing the steady state constraint $\oint \bar{\mathcal{J}}_{h}(\theta) d\cos\theta = 0$, and following the same procedure as in the previous section (with $k_i = k_i^{(0)} + k_i^{(1)} \cos \theta$), we obtain the same expression for the proton flux as equation (\ref{H:flux:approximate})
\begin{equation} 
\bar{\mathcal{J}}_h(\theta)  \cong \left[ \Delta k_{\text{eff}}^{(h)} \  \left( 1 - 2 \cos \theta \right)   - k_{\text{eff}}^{(h)} \ \delta \left(\Phi + C_h \right) \right] \bar{c}_{hp}^{\infty} \  K(\cos \theta)  \ ,  \label{H:flux:approximate2}
\end{equation}
where here we have 
\begin{align}
k_{\text{eff}}^{(h)} & = \frac{4 \{k_0 k_2k_3\}^{(0)}}{\mathcal{M}^{(0)} } \ ;  \qquad  
\Delta k_{\text{eff}}^{(h)}   = \frac{ k_{\text{eff}}^{(h)} }{2} \left( \frac{ k_4^{(1)}  }{ k_4^{(0)} }  - \frac{k_3^{(1)}}{k_3^{(0)}} \right) \ .  \label{gamma:1:definition2}
\end{align}
The deviation $\delta \left( \Phi + C_h \right) = \left[ ( \Phi + C_h) - \int_0^{\pi}  ( \Phi + C_h) K(\cos\theta) \ \sin \theta \ d  \theta \right]$ retains its previous definition as given in equation (\ref{eq:deviations}). Finally, we obtain the same relation  $\varphi_s  = \varphi_s^{\ominus} - \ln 10 \ \mbox{pH}$ and the kinetically defined potential for this scheme is $\varphi_s^{\ominus} = - (1/2) \ln \left( k_3^{(0)}/k_4^{(0)} \right)$.} 

\tbl{It is noteworthy that both reaction schemes possess many similar qualitative features: the solute fluxes $\bar{\mathcal{J}}_i$'s retain the same functional dependence on the fuel concentraton $\bar{c}_{hp}$ and the variation in reaction rates. }

\iffalse
The proton flux $\mathcal{J}_h$ in eqn. (\ref{proton:flux:definition}) automatically satisfy the steady state requirement (\ref{quasi:steady:constraint}) and hence the conservation of total charge on the swimmer surface. This constraint is reflected by the Nernst-equation for the Pt shell (see Appendix \ref{appA} for the derivation)
\begin{equation}
\varphi_s  \cong \varphi_s^{\ominus} - \ln 10 \ \mbox{pH} \quad ; \quad  \varphi_s^{\ominus} \cong - \frac{1}{2} \ln \left( \frac{\{ k_1 k_3\}^{(0)} }{ \{ k_0 k_{-3} \}^{(0)}} \right) \ , \label{zeta:kinetics}  
\end{equation}  
where the solution pH and the reaction equilibrium kinetics fixes the surface (average) potential $(\varphi_s)$ and hence the surface charge. $\varphi_s^{\ominus}$ is the reaction equilibrium kinetically defined standard potential of the Pt-cap.
%\yhy{The average zeta potential $\left< \zeta_0\right>= (\zeta_{\mbox{pt}}+\zeta_{\mbox{ps}})/2$ and the dipole moment $\left< \cos\theta \zeta_0\right>$ are accessible experimentally via linear and rotational electrophoresis~\citep{ebbens2015boundaries}. 
\fi

\section{Derivation of the slip velocity}\label{appB}
\yhy{In the Debye-layer, where the fields varies on the Debye-lengthscale $\kappa^{-1}$, we re-scale the radial coordinate by the $\lambda = (\kappa a)^{-1}$,
\begin{equation}
\rho = \frac{r-1}{\lambda} \ ,
\end{equation}
and expand the deviation fields in the form 
\begin{align}
%F(r,\theta) & = \mathcal{F}^{(0)}(\rho,\theta) + \lambda \ \mathcal{F}^{(1)}(\rho,\theta) + \cdots \quad (\mbox{for $C_i, \Phi, v_r$ and $v_{\theta}$})  \nonumber \\
C_i(r,\theta) & = \mathcal{C}^{(0)}_i(\rho,\theta) + \lambda \ \mathcal{C}^{(1)}_i(\rho,\theta) +\cdots  \ ,  \label{inner:expansion:Ci} \\
\Phi(r,\theta) & = \mathcal{\varphi}^{(0)}(\rho,\theta) + \lambda \ \mathcal{\varphi}^{(1)}(\rho,\theta) + \cdots   \label{inner:expansion:phi}\\
\bs{v}(r,\theta) & = \bs{\mathcal{V}}^{(0)}(\rho,\theta) + \lambda \ \bs{\mathcal{V}}^{(1)}(\rho,\theta) + \cdots \ ,   \label{inner:expansion:v}  \\
p(r,\theta) & = \lambda^{-2} \mathcal{P}^{(-2)}(\rho,\theta) + \lambda^{-1} \mathcal{P}^{(-1)}(\rho,\theta) + \cdots 
%\quad \rho = \frac{r-1}{\lambda} 
\end{align}
where $i \in \{ h,oh,s\pm \}$.
It is noteworthy that the expansion for the pressure field begins with $\mathcal{P}^{(-2)}$ to balance $\mathcal{O}\left(\lambda^{-2} \right)$ radial electric stresses that could not be accounted by the viscous stresses at the interface~\citep{Anderson1989,Yariv2011}. \\
}  \\

\subsection{Ionic solute concentrations}
\tl{ 
Exploiting the axisymmetry of the problem, we write the steady state Nernst-Planck equations (\ref{ions:nernstplanck}) in spherical polar coordinates, with only radial and polar angle dependence. 
\begin{align}
\nabla \cdot \mathbf{J}_i (r,\theta) = \left( \frac{\partial }{\partial r} + \frac{2}{r} \right) J_{i,r} + \frac{1}{r}\left( \frac{\partial }{\partial \theta} + \cot \theta \right) J_{i,\theta} = 0 \ .    \label{mass:trans:eqns}
\end{align} 
where
\beqa
{J}_{i,\theta}  &=& - {r}^{-1} \partial_\theta C_{i}   -  z_i ( 1 + C_{i} ) r^{-1} \partial_\theta \Phi   \\
{J}_{i,r}  &=& - \partial_r C_{i}   -  z_i ( 1 + C_{i} )\partial_r \Phi   \ ,
\eeqa
}

\yhy{ 
%In this inner (Debye) layer, we introduce a stretched coordinate, $\rho = (r-1)/\lambda$, and 
We therefore expand the fluxes in the inner coordinates $(\rho,\theta)$, noting that $r = \lambda \rho + 1$
%Therefore, to leading order in the inner expansion, we write 
\begin{equation}
\mathbf{J}_i(r,\theta) \ = \  \bs{\mathfrak{J}}_i(\rho,\theta) = \  \lambda^{-1} \ \bs{\mathfrak{J}}_i^{(-1)}(\rho,\theta)  \  +   \   \bs{\mathfrak{J}}_i^{(0)}(\rho,\theta) \  +  \  \mathcal{O}(\lambda)  \   ,  
\end{equation}
where we define radial and \tl{polar} components of the currents (see eqns. \ref{inner:expansion:Ci} and \ref{inner:expansion:phi})
\begin{align}
\mathfrak{J}^{(-1)}_{i,\theta}(\rho,\theta)  &  =   \  0  \   , \\
\mathfrak{J}^{(-1)}_{i,\rho}(\rho,\theta) & =  \  - \frac{\partial \mathcal{C}^{(0)}_i}{ \partial \rho } - z_i \left(1  + \mathcal{C}^{(0)}_i \right) \frac{\partial \varphi^{(0)}}{ \partial \rho }  \ , \label{eq:j-1} \\
\mathfrak{J}^{(0)}_{i,\theta}(\rho,\theta)   & = \  - \frac{\partial \mathcal{C}^{(0)}_i}{ \partial \theta } - z_i \left(1  + \mathcal{C}^{(0)}_i \right) \frac{\partial \varphi^{(0)}}{ \partial \theta }  \ ,   \\
 \mathfrak{J}^{(0)}_{i,\rho}(\rho,\theta) & = \  -  \frac{\partial \mathcal{C}^{(1)}_i}{ \partial \rho } - z_i \frac{\partial \varphi^{(1)}}{ \partial \rho } - z_i \mathcal{C}^{(1)}_i \frac{\partial \varphi^{(0)}}{ \partial \rho }  \  .
\end{align}
Hence,  equation (\ref{mass:trans:eqns}) can be written
%the above Nernst-Planck equations, with $\mathbf{J}_i(r,\theta) = \bs{\mathfrak{J}}_i(\rho,\theta)$, 
\begin{align}
\nabla \cdot \bs{\mathfrak{J}}_i (\rho,\theta) & =  \frac{1}{\lambda}\frac{\partial \mathfrak{J}_{i,\rho} }{\partial \rho} + \frac{2}{(1 + \lambda \rho)} \mathfrak{J}_{i,\rho} + \frac{1}{(1 + \lambda \rho)}\left( \frac{\partial }{\partial \theta} + \cot \theta \right) \mathfrak{J}_{i,\theta}   \  ,
\end{align}
%where $J_{i,\rho}$ and $J_{i,\theta}$ are the radial and azimuthal components of the solute currents $\mathbf{J}_i$.
%\begin{align}
%\mathbf{J}_i (\rho,\theta) & = - \ \frac{\bs{\hat{e}}_{\rho}}{\lambda} \ \left( \frac{\partial C_i }{\partial \rho} + z_i \left( 1 + C_i \right) \frac{\partial \Phi}{\partial \rho}  \right) \nonumber \\ & \ \quad - \frac{\bs{\hat{e}}_{\theta}}{\left( 1 + \lambda \rho \right)} \ \left( \frac{\partial C_i }{\partial \theta} + z_i \left( 1 + C_i \right) \frac{\partial \Phi}{\partial \theta}  \right) \\
%\nabla \cdot \mathbf{J}_i (\rho,\theta) & =  \frac{1}{\lambda}\frac{\partial J_{i,\rho} }{\partial \rho} + \frac{2}{(1 + \lambda \rho)} J_{i,\rho} + \frac{1}{(1 + \lambda \rho)}\left( \frac{\partial }{\partial \theta} + \cot \theta \right) J_{i,\theta} 
%\end{align}
from which performing an expansion in $\lambda$ and equating terms order by order gives the following equations at order 
%$\lambda^n$
\begin{align}
\lambda^{-2}:& \qquad \frac{\partial \mathfrak{J}^{(-1)}_{i,\rho}}{\partial \rho} = 0  \ ,   \qquad   \Rightarrow  \quad   \mathfrak{J}^{(-1)}_{i,\rho}(\rho,\theta)  \ = \  {\mathfrak h}_i(\theta)  \ ,    \label{flux:equation2} \\ 
\lambda^{-1}:&  \qquad \frac{\partial \mathfrak{J}^{(0)}_{i,\rho}}{\partial \rho} + 2 \mathfrak{J}^{(-1)}_{i,\rho} = 0  \ ,   \qquad  \Rightarrow  \quad  \mathfrak{J}^{(0)}_{i,\rho} (\rho,\theta)  \ =  \   g_i(\theta)   + 2   {\mathfrak h}_i(\theta) \ \rho \  , \label{flux:equation1}   \\
\lambda^{0} :& \qquad \frac{\partial \mathfrak{J}^{(1)}_{i,\rho}}{\partial \rho} + 2 \mathfrak{J}^{(0)}_{i,\rho} - 2 \rho \mathfrak{J}^{(-1)}_{i,\rho} + \left( \frac{\partial}{\partial \theta} + \cot \theta \right) \left( \mathfrak{J}^{(0)}_{i,\theta} - \rho \  \mathfrak{J}^{(-1)}_{i,\rho}\right) =0 \label{flux:equation0}  \ , 
\end{align}
where ${\mathfrak h}_i(\theta), \  g_i(\theta)$ are arbitrary functions of $\theta$.
Matching the currents in the inner and outer regions, 
\begin{equation}
 {\mathfrak h}_i(\theta) \   =  \   \lim_{\lambda \rightarrow 0 , \  \rho \rightarrow \infty}  \mathfrak{J}^{(-1)}_{i,\rho}(\rho, \theta)  =  \lim_{r \rightarrow 1, \ \lambda \rightarrow 0}  \hat{\bs{n}} \cdot \mathbf{J}_i^{(-1)}(r,\theta)  =  0   \  , \label{eq:matchj}
\end{equation}
which implies ${\mathfrak h}_i(\theta) = 0$ for all the species. Furthermore, the next order matching 
\begin{equation}
 g_i(\theta)  \ =  \   \mathfrak{J}^{(0)}_{i,\rho}(\rho =0,\theta)   \    =  \   \lim_{\lambda \rightarrow 0 , \  \rho \rightarrow \infty}  \mathfrak{J}^{(0)}_{i,\rho}(\rho, \theta)  =  \lim_{r \rightarrow 1, \ \lambda \rightarrow 0}  \hat{\bs{n}} \cdot \mathbf{J}_i^{(0)}(r,\theta)    \  , 
\end{equation}
providing the solution $g_i(\theta) = \mathcal{J}_i(\theta) K(\cos\theta) $; where $\mathcal{J}_i(\theta)$ are defined in Appendix  \ref{appA} and $K(\cos\theta)$ is defined in equation (\ref{eq:kcos}). Therefore,  the outer flux boundary conditions at  leading order are,
\begin{equation}
\hat{\bs{n}} \cdot \mathbf{J}_i(r=1,\theta) \ =  \ g_i(\theta) = \begin{dcases}
 \mathcal{J}_h(\theta)  K(\cos\theta)  \  ;  &   i = h \ , \ \text{(protons)}  \ ,  \\
 0 \   ;  &  i \in \{ oh, s\pm\}
 \end{dcases} \  .
\end{equation}
In the following we drop the $^{(0)}$ subscript for the outer fluxes as we are interested only in the leading order contributions (i.e we have set $\mathbf{J}_i^{(n)} = 0$ for $n \geq 1$).}

\tl{Next, we obtain the concentration profiles by first matching the inner fields with the $\mathcal{O} \left( 1\right)$ outer fields $C_i(r,\theta), \Phi(r,\theta)$;}
\begin{equation}
\lim_{\lambda \rightarrow 0; \ \rho \rightarrow \infty} \{ \mathcal{C}^{(0)}_i , \varphi^{(0)} \} \left( \rho, \theta \right) \ \ = \quad  \lim_{r \rightarrow 1; \lambda \rightarrow 0} \{  C_i^{(0)}, \Phi^{(0)} \} (r,\theta) \quad  = \ \  \{ C_i, \Phi \} (1,\theta)  \ . \label{eq:matchc}
\end{equation} 
\tl{Integrating equation ({\ref{eq:j-1}) and using equations (\ref{flux:equation2},\ref{eq:matchj}) and (\ref{eq:matchc}), 
we obtain the leading order concentration profile} 
\begin{equation}
1  + \mathcal{C}^{(0)}_i(\rho,\theta) = \left( 1 + C_i(1,\theta) \right) e^{- \left( \varphi^{(0)} (\rho,\theta) - \Phi(1,\theta) \right)}  \ . \label{inner:outer:relation}
\end{equation}}
\tl{This method can be iterated to obtain the higher order concentration fields such as $\mathcal{C}^{(1)}(\rho,\theta)$ from equation (\ref{flux:equation0}) above.}

\subsection{Electric field}
 Writing out Poisson's equation, in  spherical polar coordinates, 
\begin{equation}
 - \lambda^2 \left( \frac{1}{r^2} \frac{\partial}{\partial r} r^2 \frac{\partial}{\partial r} + \frac{1}{r^2 \sin \theta} \frac{\partial}{\partial \theta} \sin \theta \frac{\partial }{\partial \theta} \right) \Phi(r,\theta) = \sum_{i \in \{ h,oh,s\pm \}}  \mathcal{Z}_i C_i  \ ,
\end{equation}
which at the leading order in the inner expansion, reduces to 
\begin{equation}
-\frac{\partial^2 \varphi^{(0)}}{\partial \rho^2} =  \sum_i \mathcal{Z}_i \mathcal{C}_i^{(0)} \ .  \label{Poisson:B:inner}
\end{equation}

\tl{Substituting the concentration profiles from eqns. (\ref{inner:outer:relation}) into to the foregoing eqn. (\ref{Poisson:B:inner}), we have 
\begin{align}
- \frac{\partial^2 \varphi^{(0)}}{\partial \rho^2}  = \quad \sum_{i \in \{h,oh,s\pm\}} \mathcal{Z}_i \Big( 1 + C_i (1,\theta) \Big) e^{- z_i \left( \varphi^{(0)}(\rho,\theta) - \Phi(1,\theta) \right)} \ .
\end{align} 
%where the matching conditions
%\begin{equation}
%\lim_{\lambda \rightarrow 0; \ \rho \rightarrow \infty} \{ \mathcal{ \mathcal{C}}^{(0)}_i  , \varphi^{(0)} \} \left( \rho, \theta \right) = \lim_{r \rightarrow 1^+} \{ C_i, \Phi \} \left( r,\theta \right) = \{ C_i,\Phi \} (1,\theta) \ ,
%\end{equation}
%was used to eliminate $\mathcal{C}^{(0)}(\rho \rightarrow \infty,\theta)$ fields. 
In addition, applying electroneutrality  in the outer region (see eqn. \ref{outer:poisson:NP1} in the main text) at  leading order,
\begin{equation}
\sum_{i \in \{h,oh,s\pm \}} \mathcal{Z}_i C_i = 0  \ ,
\end{equation}
leads to the simpler expression 
\begin{equation}
\frac{\partial^2 \varphi^{(0)}}{\partial \rho^2}  =  \yhy{C_s^*}(\theta) \ \sinh \left(\varphi^{(0)}(\rho, \theta) - \Phi(1,\theta)  \right) \ ,  \label{PB:appA}
\end{equation}
where we have defined  
\begin{equation}
\yhy{C_s^*}(\theta) = 2 \sum_{i \in \{h,s+\} } \mathcal{Z}_i \left( 1 + C_i(1,\theta) \right) \ .
\end{equation}
Introducing a convenient factor 
\begin{align}
 2 \frac{\partial \varphi^{(0)}}{\partial \rho} \frac{\partial^2 \varphi^{(0)}}{\partial \rho^2}  = 2 \yhy{C_s^*}\frac{\partial \varphi^{(0)}}{\partial \rho}   \sinh \left(\varphi^{(0)}(\rho, \theta) - \Phi(1,\theta)  \right) 
\end{align}
and integrating once gives
\begin{equation}
\left( \frac{\partial \varphi^{(0)}}{\partial \rho} \right)^2 = 2 \yhy{C_s^*}\left[ \cosh\left( \varphi^{(0)}(\rho, \theta) - \Phi(1,\theta) \right) - 1 \right]  \ ,
\end{equation}
where the matching condition $\partial_{\rho} \varphi^{(0)}(\rho \rightarrow \infty,\theta) = 0$ (since the outer electric field expansion begins at $\mathcal{O}(1)$) was applied. Thus, we obtain the electric field in the Debye-layer using the identity $\left( 2 \sinh^2(x/2) = \cosh(x) - 1 \right)$,
\begin{equation}
-\frac{\partial \varphi^{(0)}}{\partial \rho} =  2 \sqrt{ \yhy{C_s^*}} \sinh\left( \frac{\varphi^{(0)}(\rho,\theta) - \Phi(1,\theta)}{2} \right) \ .
\end{equation}
Integrating once again, we obtain
\begin{equation}
\frac{1}{2}\ln \left[ \frac{\cosh\left( \left[\varphi^{(0)}(\rho',\theta) - \Phi(1,\theta)\right]/2\right) - 1 }{ \cosh\left(  \left[\varphi^{(0)}(\rho',\theta) - \Phi(1,\theta)\right]/2 \right) + 1} \right]_{\rho' = 0}^{\rho'=\rho} = - \sqrt{ \yhy{C_s^*}}  \ \rho   \  .
\end{equation}
Now, using the hyperbolic identities $2 \begin{Bmatrix}
\sinh \\ 
\cosh \end{Bmatrix}^2 \left( \frac{x}{2} \right)
  = \cosh (x) \mp 1$, we obtain~\citep{Anderson1989,Yariv2011}
\begin{equation}
\tanh\left(\frac{\varphi^{(0)}(\rho,\theta) - \Phi(1,\theta)}{4} \right) = \tanh\left( \frac{\zeta(\theta)}{4}  \right) e^{- \sqrt{ \yhy{C_s^*}} \ \rho } \ , \label{electrostatic:potential:inner}
\end{equation}
where $ \zeta(\theta) = \varphi_s - \Phi(1,\theta)$.}
 \\

\subsection{Momentum conservation}
Writing the Stokes equations in spherical polar coordinates, 
\begin{align}
\left( \nabla \cdot {\bs\Pi} \right) \cdot \bs{\hat{e}}_r & = \nabla^2 v_r - \frac{2 v_r}{r^2} - \frac{2}{r^2 \sin \theta} \frac{\partial }{\partial \theta } \left( \sin \theta \, v_{\theta} \right) - \frac{\partial p }{ \partial r} +  \frac{\partial \Phi}{\partial r} \nabla^2 \Phi = 0 \ , \\
\left( \nabla \cdot \bs{\Pi} \right) \cdot \bs{\hat{e}}_{\theta} & = \nabla^2 v_{\theta} - \frac{ v_{\theta}}{r^2 \sin^2 \theta} + \frac{2}{r^2} \frac{\partial v_r}{\partial \theta} - \frac{1}{r} \frac{\partial p}{ \partial \theta} + \frac{1}{r}  \frac{\partial \Phi}{\partial \theta} \nabla^2 \Phi = 0  \ ,
\end{align}
%and the incompressibility condition 
\begin{equation}
\nabla \cdot \bs{v} = \frac{\partial v_r }{\partial r} + \frac{2 v_r}{r}  + \frac{1}{r \sin \theta } \frac{\partial }{ \partial \theta} \left( \sin \theta \,  v_{\theta} \right) = 0  \ .
\end{equation}
To the leading order in the inner expansion (see eqns. \ref{inner:expansion:Ci}-\ref{inner:expansion:v}), the static pressure balances the electrostatic stresses normal to the surface~\citep{Anderson1989,Yariv2011}
\begin{equation}
\bs{\hat{e}}_{\rho} : \qquad  \lambda^{-3} \left( - \frac{\partial \mathcal{P}^{(-2)}}{\partial \rho} - \sum_i \mathcal{Z}_i \mathcal{C}_i^{(0)} \frac{\partial \varphi^{(0)}}{\partial \rho} \right) + \lambda^{-2} \frac{\partial^2 \mathcal{V}^{(0)}_{\rho}}{\partial \rho^2} + \mathcal{O}\left( \lambda^{-1} \right) = 0  \ .\label{momt:bal:radial}
\end{equation}
\tl{Note that the expansion for the pressure field begins at $\mathcal{P}^{(-2)}$ to balance $\mathcal{O}\left(\lambda^{-2} \right)$ radial electric stresses that cannot be accounted by the viscous stresses~\citep{Anderson1989,Yariv2011}.}
The viscous stresses balances the static pressure gradient and tangential electric stresses along the surface 
\begin{align}
\bs{\hat{e}}_{\theta}: \qquad \lambda^{-2} \left( \frac{\partial^2 \mathcal{V}_{\theta}^{(0)}}{\partial \rho^2} - \frac{\partial \mathcal{P}^{(-2)}}{\partial \theta} - \sum_i \mathcal{Z}_i \mathcal{C}_i^{(0)} \frac{\partial \varphi^{(0)}}{\partial \theta} \right) + \mathcal{O}\left( \lambda^{-1} \right) = 0 \ , \label{momt:bal:tangential}
\end{align}
with the leading order incompressibility constraint
\begin{equation}
\lambda^{-1} \frac{\partial \mathcal{V}^{(0)}_{\rho}}{\partial \rho} + \mathcal{O} \left( 1\right) = 0   \ .   \label{incompressibility:dl}
\end{equation}
\yhy{Therefore, to leading $\mathcal{O}(\lambda^{-3})$ in eqn. (\ref{momt:bal:radial}), the static pressure balances the radial electrostatic stresses
\begin{equation}
- \frac{\partial \mathcal{P}^{(-2)}}{\partial \rho} - \sum_i \mathcal{Z}_i \mathcal{C}_i^{(0)} \frac{\partial \varphi^{(0)}}{\partial \rho} = 0    \ , \label{momentum:radial:balance}
\end{equation}
which gives the static pressure field 
\begin{equation}
\mathcal{P}^{(-2)}\left( \rho, \theta\right) = 2 \yhy{C_s^*(\theta)}\ \sinh^2 \left( \frac{\varphi^{(0)}(\rho, \theta) - \Phi(1,\theta)}{2} \right) \ , \label{Pressure:field:inner}
\end{equation}
where matching with the outer solution implies $\mathcal{P}^{(-2)}(\infty,\theta) = 0$ (since the outer field $p$ expansion begins at $\mathcal{O}(1)$).  The next order $\mathcal{O}(\lambda^{-2})$ in eqn. (\ref{momt:bal:radial}) momentum balance is 
\begin{equation}
 \frac{\partial^2 \mathcal{V}_{\rho}^{(0)}}{ \partial \rho^2} = 0 \  ,
\end{equation}
where the continuity equation (\ref{incompressibility:dl}), $\partial_{\rho} \mathcal{V}_{\rho}^{(0)} = 0$ (i.e $\mathcal{V}_{\rho}^{(0)} $ is $\rho$ independent), implies
\begin{equation}
\mathcal{V}_{\rho}^{(0)} \left( \rho, \theta\right) = \mathbf{U}^e \cdot \bs{\hat{e}}_{\rho}   \  .
\end{equation}
At $\mathcal{O}(\lambda^{-2})$ in eqn. (\ref{momt:bal:tangential}),  viscous stresses balance the tangential pressure gradient $\partial_{\theta}\mathcal{P}^{(-2)}$ and the tangential electrical stress; }
\begin{equation}
\frac{\partial^2 \mathcal{V}_{\theta}^{(0)}}{\partial \rho^2} - \frac{\partial \mathcal{P}^{(-2)}}{\partial \theta} - \sum_i \mathcal{Z}_i \mathcal{C}_i^{(0)} \frac{\partial \varphi^{(0)}}{\partial \theta} = 0  \ .
\end{equation}
Using equations (\ref{Pressure:field:inner}) and (\ref{PB:appA}), we obtain
\begin{align}
\frac{\partial^2 \mathcal{V}_{\theta}^{(0)}}{\partial \rho^2}  & = 2 \sinh^2 \left( 2\varphi \right) \ \frac{\partial \yhy{C_s^*}}{\partial \theta}  - 2 \yhy{C_s^*} \ \sinh \left( 2\varphi\right) \cosh\left(2\varphi \right) \  \nonumber \\
& =   \frac{ 2\tanh\left( 2\varphi \right)}{ 1 - \tanh^2\left( 2\varphi \right)} \left[ \tanh\left( 2\varphi \right) \frac{\partial \yhy{C_s^*}}{\partial \theta} + \yhy{C_s^*} \frac{\partial \Phi}{\partial \theta}  \right]  \ ,
\end{align}
where \textcolor{black}{ $4\varphi (\rho,\theta) = \varphi^{(0)}(\rho,\theta) - \Phi(1,\theta) $}. Using the identity $\tanh(2x) = 2\tanh(x)/(1 + \tanh^2(x))$, 
\begin{align}
\frac{\partial^2 \mathcal{V}_{\theta}^{(0)}}{\partial \rho^2}  & = 2 \left[ \frac{4 \tanh^2(\varphi) \ \frac{\partial \yhy{C_s^*}}{\partial \theta} }{\left( 1 + \tanh(\varphi) \right)^2 \left(1  - \tanh(\varphi) \right)^2}  -  \frac{2 \tanh(\varphi) \ \yhy{C_s^*}\frac{\partial \Phi}{\partial \theta}}{\left( 1 + \tanh(\varphi) \right)^2 \left(1  - \tanh(\varphi) \right)^2} \right]  \ . \label{stress:gradient}
\end{align} 
\yhy{It is helpful to write the coefficients of the RHS first and second terms as }
\begin{align}
\frac{4 \tanh^2(\varphi)  }{\left( 1 + \tanh(\varphi) \right)^2 \left(1  - \tanh(\varphi) \right)^2} & = \frac{-1}{\left( 1 + \tanh(\varphi) \right)} + \frac{1}{\left( 1 + \tanh(\varphi) \right)^2} \nonumber \\ & \ +  \frac{-1}{\left( 1 - \tanh(\varphi) \right)} + \frac{1}{\left( 1 - \tanh(\varphi) \right)^2} \ ,   \label{integral:identities0} \\
\frac{2 \tanh(\varphi)}{\left( 1 + \tanh(\varphi) \right)^2 \left(1  - \tanh(\varphi) \right)^2} & = \frac{1}{\left( 1 + \tanh(\varphi) \right)} + \frac{-1}{\left( 1 + \tanh(\varphi) \right)^2} \nonumber \\ & \ +  \frac{-1}{\left( 1 - \tanh(\varphi) \right)} + \frac{1}{\left( 1 - \tanh(\varphi) \right)^2}  \ . 
\end{align}
\tl{Using  eqn. (\ref{electrostatic:potential:inner})} and integrating once, we obtain 
\begin{align}
\frac{\partial \mathcal{V}_{\theta}^{(0)}}{\partial \rho} = \frac{2}{\sqrt{\yhy{C_s^*}}}  \left[ \frac{2 Q^2}{\left( Q^2 - e^{2  \sqrt{\yhy{C_s^*}} \ \rho} \right)} \frac{\partial \mathcal{C}^*}{ \partial \theta} - \frac{2Q e^{\sqrt{\yhy{C_s^*}} \ \rho}}{\left( Q^2 - e^{2  \sqrt{\yhy{C_s^*}} \ \rho} \right)} \yhy{C_s^*}\frac{\partial \Phi}{\partial \theta}\right] \ .
\end{align}
where $Q = \tanh \left( \zeta(\theta)/4 \right)$ and such that matching with the outer solution imposes $\partial_{\rho} \mathcal{V}_{\theta}^{(0)}(\infty,\theta) = 0$. \yhy{In obtaining the above expression, we have used the following integral identities}
\begin{align}
\int{ \frac{d \rho' }{1 + Q e^{- \sqrt{\yhy{C_s^*}} \ \rho'}}} & = \frac{1}{\sqrt{\yhy{C_s^*}}} \ln \left( Q + e^{{\sqrt{\yhy{C_s^*}}} \ \rho} \right) + \mbox{constant} \ ,\\ 
\int{ \frac{d \rho' }{ \left( 1 + Q e^{- \sqrt{\yhy{C_s^*}} \ \rho'} \right)^2}} & = \frac{1}{\sqrt{\yhy{C_s^*}}} \frac{Q}{ \left(  Q + e^{\sqrt{\yhy{C_s^*}}} \right)} +  \frac{1}{\sqrt{\yhy{C_s^*}}} \ln \left( Q + e^{{\sqrt{\yhy{C_s^*}} } \ \rho} \right) + \mbox{constant} \ ,\\
\int{ \frac{d \rho' }{1 - Q e^{- \sqrt{\yhy{C_s^*}} \ \rho'}}} & = \frac{1}{\sqrt{\yhy{C_s^*}}} \ln \left( e^{{\sqrt{\yhy{C_s^*}}} \ \rho} - Q \right) + \mbox{constant}  \ ,\\ 
\int{ \frac{d \rho' }{ \left( 1 - Q e^{- \sqrt{\yhy{C_s^*}} \ \rho'} \right)^2}} & = - \frac{1}{\sqrt{\yhy{C_s^*}}} \frac{Q}{ \left(  Q + e^{\sqrt{\yhy{C_s^*}}} \right)}  + \frac{1}{\sqrt{\yhy{C_s^*}}} \ln \left( e^{{\sqrt{\yhy{C_s^*}} } \ \rho} - Q \right) + \mbox{constant} \ . \label{integral:identities1}
\end{align}
Finally, integrating  again,
\begin{align}
\mathcal{V}_{\theta}^{(0)} \left( \rho,\theta \right) & = \mathcal{V}_{\theta}^{(0)}(0,\theta) + \frac{2}{\yhy{C_s^*}}  \frac{\partial \yhy{C_s^*}}{\partial \theta} \ln\left( \frac{1 - Q^2 e^{ - 2 \rho  \sqrt{\yhy{C_s^*}}}}{1 - Q^2} \right)  \nonumber \\ & \qquad \qquad - 4 \frac{\partial \Phi}{\partial \theta} \left( \tanh^{-1}\left( Q e^{- \sqrt{\yhy{C_s^*}} \rho}\right) - \tanh^{-1}\left( Q\right)\right) \ . \label{eq:inner_flow}
\end{align} 
\yhy{Therefore, in the thin-layer limit $\lambda \rightarrow 0$ and $\rho \rightarrow \infty$,} 
\begin{align}
\lim_{\lambda \rightarrow 0; \ \rho \rightarrow \infty} \mathcal{V}_{\theta}^{(0)} \left( \rho ,\theta \right) & = \mathcal{V}_{\theta}^{(0)}(0,\theta) +  \zeta(\theta) \frac{\partial \Phi}{\partial \theta} - \frac{2}{\yhy{C_s^*}}  \frac{\partial \yhy{C_s^*}}{\partial \theta} \ln\left( 1 - \tanh^2\left( \frac{\zeta(\theta)}{4}\right)\right)     \ .
\end{align}
\yhy{Finally, matching with the leading order outer flow field
\begin{equation}
\lim_{\lambda \rightarrow 0; \ \rho \rightarrow \infty} \mathcal{V}_{\theta}^{(0)} \left( \rho ,\theta \right)  \ =  \   \lim_{r \rightarrow 1, \  \lambda \rightarrow 0}  \bs{v}^{(0)}(r,\theta)  \   =   \    \bs{v}(1,\theta)  \ , 
\end{equation}
we obtain the slip velocity boundary condition for the outer flow~\citep{AndersonPrieve1982,Yariv2011} }
\begin{equation}
\bs{v}(1,\theta) = \mathbf{U}^e  +  \left[   \zeta(\theta) \frac{\partial \Phi}{\partial \theta} + 4  \frac{\partial \ln \yhy{C_s^*}}{\partial \theta} \ln \cosh \left(\frac{\zeta(\theta)}{4} \right) \right] \bs{\hat{e}}_{\theta} \ .
\end{equation}

%\section{List of symbols}
%
%\begin{tabular}{rl}
%$a$ & Swimmer radius (size) \\
%$\mathbf{U}$ & Self-electrophoretic contribution to propulsion velocity\\
%$\mathbf{U}^d$ & Self-diffusiophoretic contribution to propulsion velocity\\
%$\mathbf{U}^{\text{tot}}$ & Propulsion velocity (combined)
%\end{tabular}

\bibliographystyle{jfm}
%% Note the spaces between the initials
%
%
%\bibliography{self_phoresis}
\bibliography{library_old,library,self_phoresis}

\end{document}